\documentclass[onecolumn, trackchanges]{aastex701}

\usepackage{amsmath,amstext}
\usepackage[T1]{fontenc}
\usepackage[figure,figure*]{hypcap}
\usepackage{comment}
\usepackage{graphicx}
\usepackage{natbib}
\usepackage{epsfig}
\usepackage{epstopdf}
\usepackage{enumerate}
\usepackage{tabularx}
\usepackage{multirow}
\usepackage{booktabs}
\usepackage{color}
\usepackage[normalem]{ulem}  
\usepackage{xspace}
\usepackage{tabstackengine}
\usepackage{xparse}

\graphicspath{{./}{Figures/}}

\accepted{by ApJ}

\newcommand{\pd}[3][]{\frac{\partial^{#1} #2}{\partial #3^{#1}}}
\newcommand{\msun}{M$_{\odot}$\xspace}
\newcommand{\eff}{\mathrm{eff}}

\newcommand{\turb}{\mathrm{turb}}
\newcommand{\postb}{\mathrm{pb}}
\newcommand{\gain}{\mathrm{gain}}
\newcommand{\thermal}{\mathrm{thermal}}

\allowdisplaybreaks 

\begin{document}

\title{Turbulence in Core-Collapse Supernovae}
\shorttitle{Turbulence in Supernovae}

\author[0009-0000-5846-5984]{David Calvert} 
\email{djcalve2@ncsu.edu}
\affiliation{Department of Physics, North Carolina State University, Raleigh, NC 27606 USA}

\author[0000-0002-4099-2978]{Michael Redle}
\email{mtredle@ncsu.edu}
\affiliation{Department of Mathematics, North Carolina State University, Raleigh, NC 27606 USA}
\affiliation{Institute of Applied and Computational Mathematics, RWTH Aachen University, 52062 Aachen, Germany}

\author[0000-0002-3211-3427]{Bibek Gautam}
\email{bgautam2@ncsu.edu}
\affiliation{Department of Physics, North Carolina State University, Raleigh, NC 27606 USA}

\author[0000-0002-5667-7577]{Charles J.\ Stapleford}
\email{staplefordcj@gmail.com}
\affiliation{Oak Ridge National Laboratory, Oak Ridge, TN 37830 USA}

\author[0000-0003-0191-2477]{Carla Fr\"{o}hlich}
\email{cfrohli@ncsu.edu}
\affiliation{Department of Physics, North Carolina State University, Raleigh, NC 27606 USA}

\author[0000-0002-3502-3830]{James P.\ Kneller}
\email{jpknelle@ncsu.edu}
\affiliation{Department of Physics, North Carolina State University, Raleigh, NC 27606 USA}

\author{Matthias Liebendorfer}
\email{matthias.liebendoerfer@unibas.ch}
\affiliation{University of Basel, Basel, Switzerland}

\shortauthors{Calvert \emph{et al.}}
\correspondingauthor{Carla Fr\"{o}hlich}

\begin{abstract}
It is understood in a general sense that turbulent fluid motion below the shock front in a core-collapse supernova stiffens the effective equation of state of the fluid and aids in the revival of the explosion. 
However, when one wishes to be precise and quantify the amount of turbulence in a supernova simulation, one immediately encounters the problem that turbulence is difficult to define and measure.
Using the 3D magnetohydrodynamic code ELEPHANT, we study how different definitions of turbulence change one's conclusions about the amount of turbulence in a supernova and the extent to which it helps the explosion. 
We find that, while all the definitions of turbulence we use lead to a qualitatively similar growth pattern over time of the turbulent kinetic energy in the gain region, the total amount of turbulent kinetic energy, and especially the ratios of turbulent to total kinetic energy, distinguish them. 
Some of the definitions appear to indicate turbulence is a necessary contributor to the explosion, and others indicate it is not. 
The different definitions also produce turbulence maps with different correlations with maps of the enstrophy, a quantity widely regarded as also indicating the presence of turbulence. 
We also compute the turbulent adiabatic index and observe that in regions of low enstrophy, this quantity is sensitive to the definition used.
As a consequence, the effective adiabatic index depends upon the method used to measure the turbulence and thus alter one's conclusions regarding the impact of turbulence within the supernova. 
\end{abstract}

\keywords{}

\section{Introduction}
\label{sec:intro}

Core-collapse supernovae (CCSNe) are the explosive deaths of stars with initial masses greater than $\sim 8-10$~\msun ~\citep{RevModPhys.74.1015,2009MNRAS.395.1409S,2015ApJ...810...34W}.
In addition to the spectacular fireworks, nuclear reactions in the ejected material will contaminate the interstellar medium of the host galaxy with freshly synthesized elements \citep{2006NuPhA.777..424N,push3,2024Univ...10..148B}.
Stellar-mass black holes, neutron stars, pulsars, and magnetars are born out of these explosions \citep{1986ApJ...307..178B,1992ApJ...392L...9D,2011Ap&SS.336..129N,2015SSRv..191..315M,2017ApJ...837..128W,2017ARA&A..55..261K,push2}, and observations indicate that Long Gamma-Ray Bursts are also associated with CCSNe \citep{1999Natur.401..453B,2006ARA&A..44..507W}. 
Despite the importance of CCSNe in understanding many aspects of the universe, the mechanism that drives CCSN explosions is not yet fully understood.
At the present time, the prevailing theory for CCSNe is based upon the `delayed neutrino-heating mechanism' \citep{1985ApJ...295...14B,1995ApJ...450..830B,2025arXiv250214836J}.
In this paradigm, an initial collapse of the stellar core forms a proto-neutron star (PNS) and launches a shock-wave into the outer layers.
The shock stalls 100-200 km above the PNS as its energy is consumed dissociating infalling nuclei and is then revived through energy deposition due to neutrino scattering / absorption in the material beneath the shock.
However simulations of supernovae in spherical symmetry fail to explode via the delayed neutrino-heating mechanism \citep{2001PhRvL..86.1935M,2002A&A...396..361R,2003ApJ...592..434T,2004ApJS..150..263L,2005ApJ...629..922S}. 
This poses a limitation for many open questions that rely on successful explosions, e.g.\ predicting supernova nucleosynthesis yields, unveiling the progenitor-remnant connection, and many more. One approach around this limitation is the use of so-called effective models. These are simulations in spherical symmetry where one aspect of multi-dimensional simulations is parametrized and the parameters are calibrated against observations. For example, P-HOTB \citep{2016ApJ...818..124E} uses a simple approximation of increased neutrino luminosities from the proto-neutron star. The PUSH method \citep{push1,push2} instead parametrizes the effectively enhanced neutrino energy-deposition in the gain layer that is seen in multi-dimensional simulations. And finally, STIR \citep{2020ApJ...890..127C} parametrizes the strength of turbulence that is observed in multi-dimensional simulations in a mixing-length type parametrization. 
Welcomely, modern \emph{multi-dimensional} simulations have achieved unaided explosions~\citep{
2007PhR...442...38J,
2017hsn..book.1095J,
2017RSPTA.37560271C,
Kuroda_2020,
2021ApJ...915...28B,
2022MNRAS.514.3941N,
2022MNRAS.510.4689V,
sykes_long-time_2024,
2024MNRAS.531.3732S,
2024ApJ...962...71W}, 
demonstrating the viability of the delayed neutrino-heating mechanism and emphasizing the importance of multi-dimensional effects in aiding explosions. 
One of these multi-dimensional effects is fluid turbulence.

As a result of the violent fluid motion in the gain region of the supernova, the low viscosity, and the significant shear stress, the fluid (or some fraction of it) can become turbulent due to instabilities.
High-resolution studies of turbulence in 3D simulations of CCSNe \citep{2011ApJ...742...74M,2012ApJ...749..142F,2013ApJ...765..110D,2013PhST..155a4022E,2014ApJ...783..125H,2015ApJ...808...70A,2015ApJ...799....5C,2016ApJ...820...76R,2020PhyS...95f4005C} are able to see the inertial range of the turbulence \citep{2016ApJ...820...76R} and the `turbulence bottleneck' -- the accumulation of power in the inertial scales of the turbulence \citep{1994PhFl....6.1411F,2003PhRvE..68b6304D,2015PhRvE..92c3009K,2015ApJ...808...70A}. 
Turbulence has the potential to serve as an aid to the shock revival by stiffening the effective adiabatic index of the fluid.
However this conclusion about the effect of turbulence is inherently dependent on how the amount of turbulence present is measured, which is a notoriously difficult problem \citep{2004iit..book.....T}. 

The purpose of this paper is to investigate how different measures of turbulence in a CCSNe alter the conclusions one draws about its effect upon the explosion.
To explore this question, we have undertaken four simulations of core-collapse supernovae spanning initial masses from 15\msun to 27\msun. 
The simulations were performed in three spatial dimensions using the ELEPHANT code~\citep{Liebendoerfer.IDSA:2009,kappeli2011fish,Cabezon-3Dcomparison}.
This code is well suited to this study because of its good spatial resolution below the shock. 
We leverage this high resolution in the gain region to explore three different methods to compute turbulence. 
Using a variety of post-processing tests, we show that the three methods produce results that are not consistent with one another, in turn presenting the strengths and potential drawbacks of each method.

Our paper is organized as follows.
The code we use, the progenitors, and general results of the simulations are described in Section \ref{sec:simulations}. We next discuss in Section \ref{sec:turbKE} the different methods we will use for measuring the turbulence, and then apply the measures to the simulations in Section \ref{sec:results}. Finally, in Section \ref{sec:TurbulentAdiabticIndex} we examine how turbulence alters the effective adiabatic index of the fluid and how this index evolves with time in the simulations.  
We summarize and present our conclusions in Section \ref{sec:conclusions}.


\section{The Simulations}
\label{sec:simulations}

\subsection{The ELEPHANT Code} \label{sec:code}

The ELEPHANT code solves the ideal hydrodynamics equations in a modified Newtonian gravitational potential together with the Isotropic Diffusion Source Approximation (IDSA) for the neutrino transport scheme. 
The code uses operator splitting between the solutions of the hydrodynamic equations and the neutrino transport equations.
In the hydrodynamics module, which has roots in the FISH code described  in~\citet{kappeli2011fish}, the following conservation laws coupled with gravitational source terms are numerically solved in three spatial dimensions:
\begin{eqnarray}
    \pd{\rho}{t}+ \boldsymbol{\nabla} \boldsymbol{\cdot} (\rho\, \boldsymbol{v} ) &= 0, \label{eq:mass}\\
    \pd{}{t}(\rho\, v_i) +
     \pd{}{x_j}(\rho \, v_i \, v_j + P_{\thermal} \,\delta_{ij}) & = -\rho\, \pd{\phi}{x_i} ,\label{eq:momentum}\\
    \pd{E}{t} + \boldsymbol{\nabla} \boldsymbol{\cdot} \left[(E+P_{\thermal})\,\boldsymbol{v} \right] & = - \rho \,\boldsymbol{v} \boldsymbol{\cdot} \boldsymbol{\nabla} \phi,\label{eq:energy}\\
    \pd{}{t}(\rho\, Y_e)+ \boldsymbol{\nabla} \boldsymbol{\cdot} (\rho \,Y_e\, \boldsymbol{v} ) & = 0 ,\label{eq:electron_fraction}\\
    \pd{}{t}(\rho\, Y_\nu^t) +\boldsymbol{\nabla} \boldsymbol{\cdot} (\rho\, Y_\nu^t\, \boldsymbol{v} ) & = 0 ,\label{eq:electron_neutrino}\\
    \pd{}{t}\left[\left(\rho\, Z_\nu^t\right)^{\frac{3}{4}}\right] + \boldsymbol{\nabla} \boldsymbol{\cdot} \left[(\rho\, Z_\nu^t)^\frac{3}{4} \boldsymbol{v} \right] & = 0,\label{eq:mean_neutrino_energy} \\
    \boldsymbol{\nabla}^2 \phi& = 4\,\pi\, G\, \rho, \label{eq:gravity}    
\end{eqnarray}
where the calculated unknowns with respect to space $x_i$ and time $t$ are the baryonic mass density $\rho$, velocity $v_i$, electron fraction $Y_e$, and the gravitational potential $\phi$. While ELEPHANT is able to include magnetic fields, the magnetic field terms have been set to zero for this study.
The temperature $T$ is not explicitly seen in these equations, it is represented in the specific internal energy $e=e(\rho,T,Y_e)$, and the total energy $E = \rho\, e + \rho\, v^2/2$. As usual, $G$ is the gravitational constant. We compute the thermal pressure $P_{\thermal} = P_{\thermal}(\rho,T, Y_e)$ using the LS220 equation of state (EOS) with an incompressibility parameter $K=220$~MeV \citep{lattimer1991generalized}.
Equations \eqref{eq:electron_neutrino}--\eqref{eq:mean_neutrino_energy} advect the trapped neutrino fractions, $Y_\nu^t$, and a multiple of the neutrino entropies, $\left(\rho Z_\nu^t\right)^{3/4}$ (where $Z_\nu$ is the mean neutrino specific energy) from the neutrino transport. 
The gravitational potential is modified following \citet{marek2006exploring} to include general relativistic corrections to the Newtonian potential.

The hydrodynamics equations \eqref{eq:mass}--\eqref{eq:gravity}  are solved using the second-order relaxation scheme of \citet{jin1995relaxation}. 
To enforce that the scheme is total variation diminishing (a nonlinear constraint to ensure stability), ELEPHANT implements the minmod limiter in supersonic flow regimes and the van Leer limiter in subsonic regimes \citep{kappeli2011fish}. 
For the time integration, a second-order predictor-corrector method is used.

For the neutrino transport, the ELEPHANT code uses the Isotropic Diffusion Source Approximation (IDSA, \citet{Liebendoerfer.IDSA:2009}). 
The underlying idea of IDSA is that the neutrinos are separated into a trapped neutrino distribution function, $f_{\nu}^{t}$, and a free-streaming neutrino distribution function, $f_{\nu}^{s}$. 
These two components are then evolved separately.
The trapped neutrino distribution functions are constructed from $Y_{\nu}^{t}$ and $Z_{\nu}^{t}$ assuming a thermal spectrum.
Then, the diffusion equation, given as,
\begin{equation}
    \frac{1}{c}\frac{\partial f_\nu^t}{\partial t} = j_\nu -(j_\nu + \chi_\nu)\,f_\nu^t - \Sigma_\nu,
    \label{eq:idsa-diff-eq}
\end{equation}
where
\begin{equation}
    \Sigma_\nu = \min \left\{\max \left[\alpha_\nu +\frac{1}{2} \, (j_\nu + \chi_\nu) \, \int f_o^s\ d\mu,\ 0\right], j_\nu \right\}, 
\end{equation}
with
\begin{equation}
    \alpha_\nu = \boldsymbol{\nabla} \boldsymbol{\cdot} \left(\frac{-1}{3\,(j_\nu + \chi_\nu+ \phi_\nu)}\ \boldsymbol{\nabla} f_\nu^t\right),
\end{equation}
is solved in three dimensions.
In these equations, $f_o^s$ denotes the use of the streaming neutrinos from the previous time step, $j_\nu$ is the spectral neutrino emissivity, $\chi_\nu$ is the neutrino absorptivity, and $\phi_\nu$ includes isoenergetic scattering in the mean free path (see \citet{bruenn1985stellar}).
The non-local value $\alpha_\nu$, a descriptor of the evolution of trapped neutrinos, is computed by explicit finite differencing, with all other unknowns solved locally using an implicit Euler time step in which the solution is found using the Newton-Raphson iterative solver. 
The numerical solution of equation \eqref{eq:idsa-diff-eq} sets the net interaction rates between the matter and neutrino radiation and updates the electron fraction $Y_e$ and the specific internal energy $e$ so that they are consistent. 
The rates $j_\nu,\ \chi_\nu$, and $\phi_\nu$ are then used to update the trapped neutrino fraction $Y_\nu^t$ and mean energy $Z_\nu^t$ seen in equations \eqref{eq:electron_neutrino} and \eqref{eq:mean_neutrino_energy}. 
This iterative method is easily parallelizable for IDSA because the only non-local value $\alpha_{\nu}$ is computed explicitly, thus implying that zones do not need to communicate with each other within the iterative solve.

ELEPHANT uses a unique combination of 3D and 1D grids as its computational domain. The innermost computational domain consists of a 3D Cartesian cuboid, and the 1D domain extends from the center of the supernova out well beyond the 3D domain and so well into the outer layers of the progenitor star. The inner 3D grid uses a uniform fixed grid spacing,  while the larger, spherically symmetric 1D domain surrounding the 3D cuboid is evolved on an adaptive grid using the Agile-IDSA solver \citep{Liebendoerfer.IDSA:2009}. 
More details of how the hydro and neutrino transport equations are solved on the 3D uniform Cartesian mesh, together with test cases, can be found in \citet{kappeli2011fish}. 
The spherically symmetric solution of the hydro and neutrino transport equations on the 1D grid is used primarily for the properties of the material falling through the boundary of the 3D mesh, but since it extends into the inner part of the Cartesian grid, it also provides a second solution of the equations that is especially useful within the PNS where it serves as a check of the entropy evolution.
In addition, the combination of the two grids with the 1D solution extending well beyond the 3D, gives ELEPHANT the ability to enlarge the 3D Cartesian cuboid if necessary. 
With this setup of the computational domain, the ELEPHANT code has high 3D spatial fidelity in the region below the shock without the burden of a large number of grid zones where the solution is spherically symmetric. Hence, it is ideally suited for the proposed study of the post bounce phase especially for the growth of the turbulence in the shocked material.
For each simulation in this study we adopt a uniform fidelity of 1~km throughout the 3D volume; initially, the 3D volume encompasses the innermost 600~km of the star (i.e., along each axis the 3D Cartesian coordinate ranges from -300~km to +300~km). 
A direct comparison of the resolution of the 3D volume with the resolution of the grids adopted elsewhere in the literature for studies of turbulence in supernovae, is challenging because of the wide variety of methods used: see, for example, the grids used in the studies by \citet{10.1093/mnras/stz2730} and \citet{2020PhyS...95f4005C}. The most similar studies to this work are by \citet{Endeve_2012,2013PhST..155a4022E} which were also Cartesian and where the highest resolution simulation used a cell spacing of $1.17\;{\rm km}$. 


\subsection{General Properties of the Simulations \label{sec:general_results}}

\begin{figure}
\centering
    \vspace{8pt}
    \includegraphics[width=0.45\textwidth]{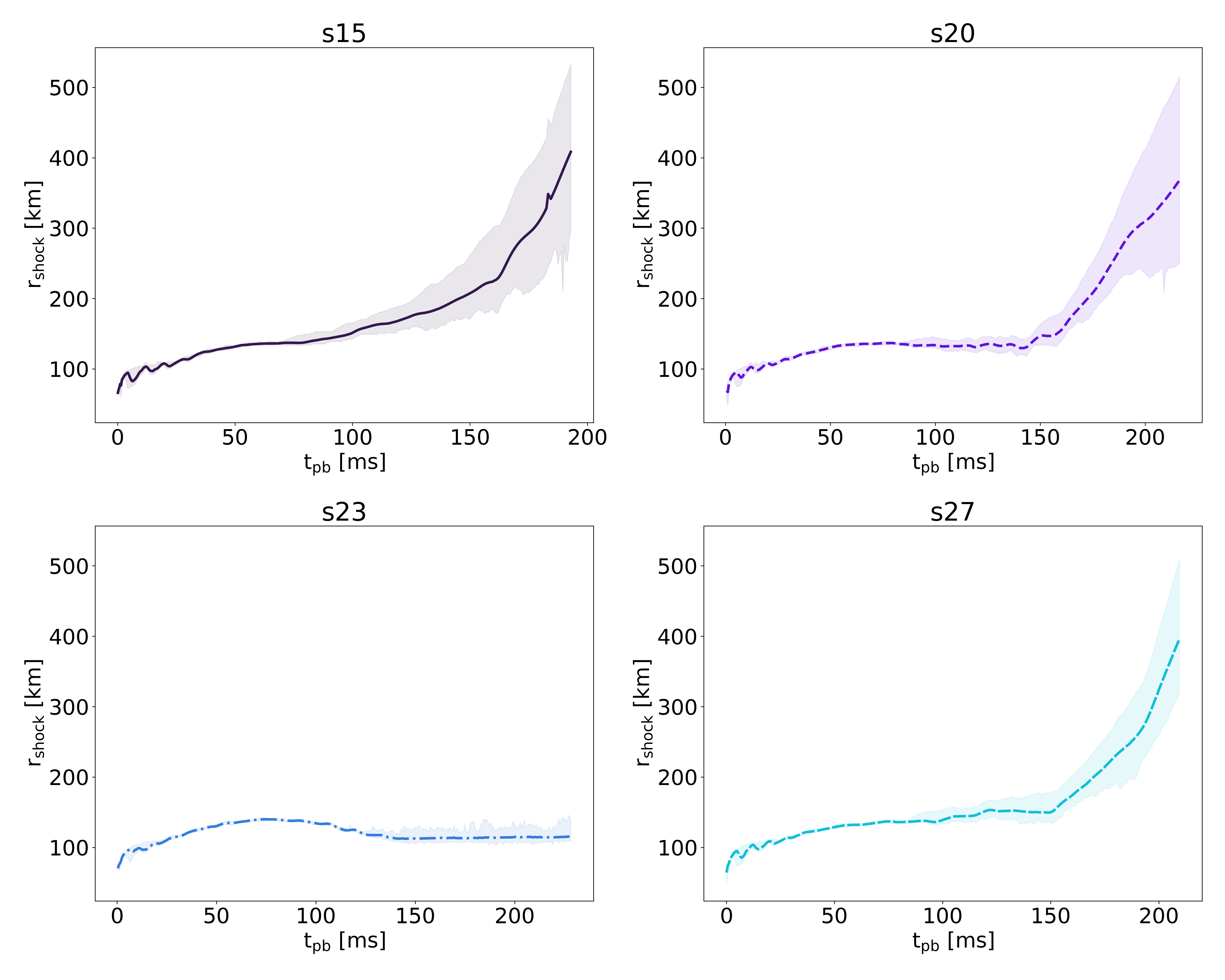}
\caption{
Average shock radii (lines), together with the minimum and maximum shock radii (shaded bands) as a function of the post-bounce time for all four simulations. The simulations end when the shock radius reaches the boundary of the largest (500 km) 3D domain. Note that the s23 simulation does not undergo a successful shock revival during the simulation time. 
\label{fig:shockradii}
}
\end{figure}

\begin{figure*}
\centering
    \includegraphics[trim = 3.9cm 0.4cm 3.0cm 0.5cm,clip,width=\textwidth]{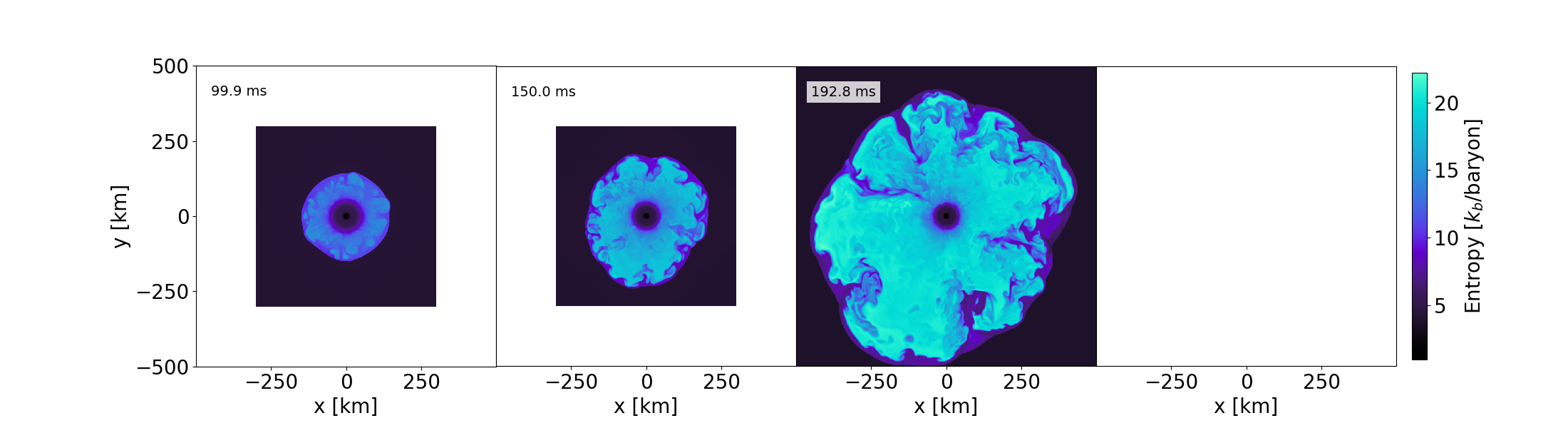} 
    
    \includegraphics[trim = 3.9cm 0.4cm 3.0cm 0.5cm,clip,width=\textwidth]{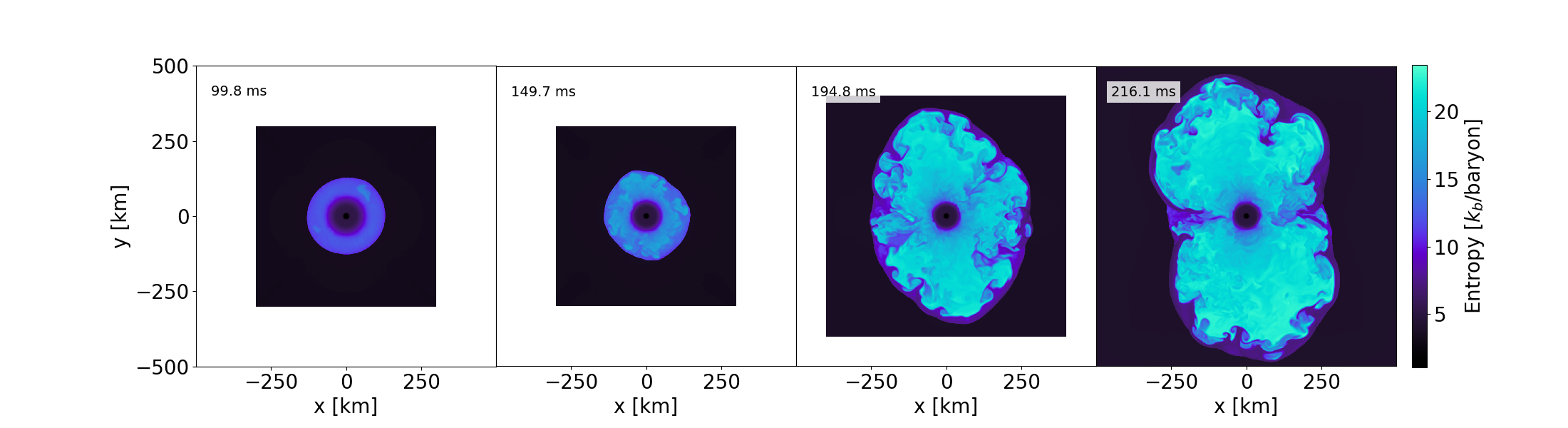} 
    
    \includegraphics[trim = 3.9cm 0.4cm 3.0cm 0.5cm,clip,width=\textwidth]{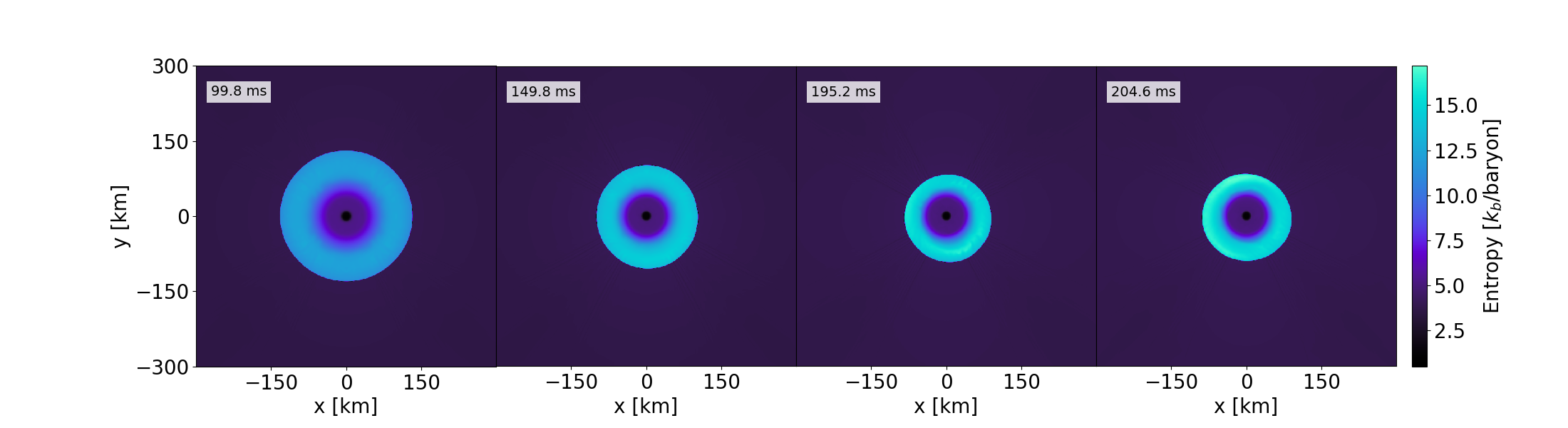} 
    
    \includegraphics[trim = 3.9cm 0.4cm 3.0cm 0.5cm,clip,width=\textwidth]{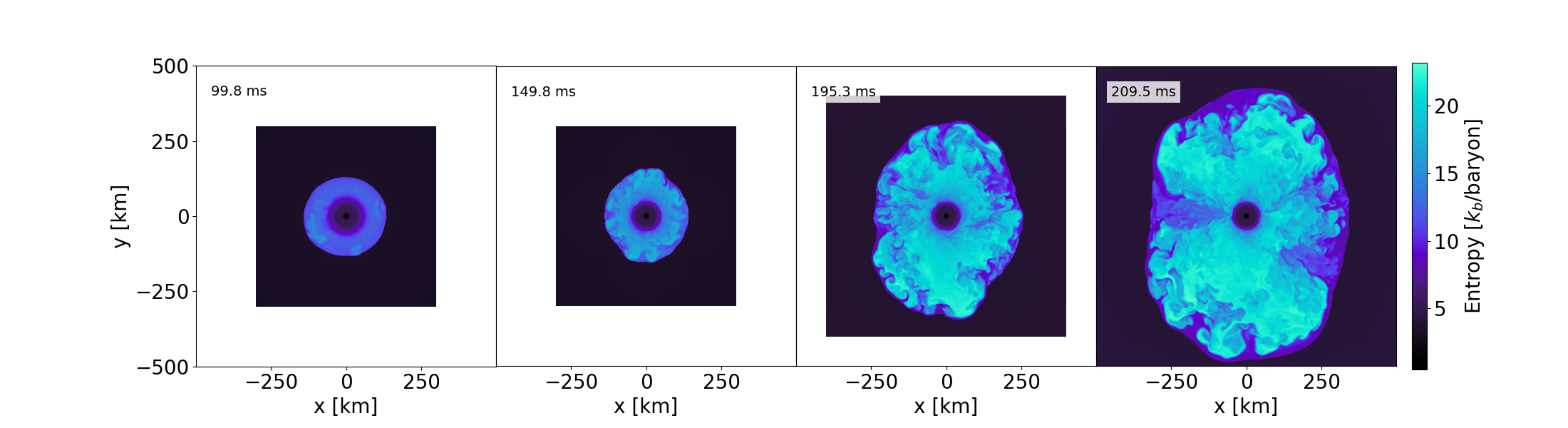} 

\caption{Entropy maps in the $xy$-plane for the s15 (top), s20 (second row), s23 (third row), and s27 (bottom row) simulations at $\sim 100$~ms (first column), $\sim 150$~ms (second column), $\sim 195$~ms (third column) post bounce, and at the final simulation time (last column). 
Note the smaller spatial scale and entropy scale for s23.
\label{fig:entropy}
}
\end{figure*}

For this study, we consider four solar-metallicity progenitor stars of 15, 20, 23 and 27 $M_{\odot}$ zero-age main sequence (ZAMS) mass \citep{woosley2007nucleosynthesis}. 
Throughout this paper we shall refer to each simulation by reference to the progenitor model using the code `sXX' where XX is the ZAMS mass in units of the solar mass.
Two of these progenitors (s15 and s20) have been extensively used in the literature. 
For the s20 model, explosions \citep{2016ApJ...818..123B,2016ApJ...825....6S} and non-explosions \citep{2016ApJ...831...81S,2018MNRAS.477.3091V} were obtained in 2D simulations. Similarly, successful and failed explosions are also reported from 3D simulations of s20, depending on the exact simulation setup and on the physics included \citep[e.g.][]{2015ApJ...808L..42M,2018ApJ...855L...3O,2019ApJ...873...45G,Kuroda_2020}. The s20 model seems to be close to the threshold between successful and failed explosions, warranting further investigations. 
For the s15 model, the picture a slightly more consistent in the sense that most simulations find successful explosion, albeit as late as $\sim 400-500$~ms post-bounce (however, see e.g.\ \citet{2016ApJ...816...43S} for a failed explosion in 2D). 
The s23 model was chosen as it is a model that often fails to explode in effective 1D simulations using the STIR or PUSH method. In a recent 3D simulation of a different 23 \msun model, the shock evolution follows other non-exploding models for the first $\sim 300$~ms post-bounce, then hovers around 170~km for another $\sim 300$~ms, and finally expands to 500~km only at 700~ms after bounce \citep{2023PhRvD.107j3015V}.
Finally, the s27 model has been used in 3D simulations where it exhibited a similar shock evolution as the s20, with shock run-away occurring at $\sim 200$~ms \citep{2018ApJ...855L...3O}. For an earlier version of the s27 (from \citet{RevModPhys.74.1015}), no shock revival was found in 3D simulations \citep{2013ApJ...770...66H,2021MNRAS.508..966T}.

As stated previously, the simultaneously computed spherically symmetric solution on the 1D grid allows ELEPHANT to expand the 3D domain if necessary. We used this feature of the code to expand the 3D domain when the maximum shock radius in the 3D solution approaches the edge of the 3D computational domain. At this point we expand the 3D domain by adding 100 grid zones in both the negative and positive directions along each coordinate axis. The new zones have the same, 1 km, resolution and the 1D solution is used to fill the fluid and neutrino variables of the new zones. We undertook this 3D domain expansion twice during the s15, s20 and s27 simulations and stopped when any Cartesian coordinate of the maximum shock position reached 500 km. This technique of expanding the 3D volume as the shock approaches the edge was also used by \citet{Endeve_2012}. 

In Figure \ref{fig:shockradii} we show the average shock radius (various line styles) for each simulation along with its minimum and maximum position (shaded bands).
During the first $t_{\postb} \approx 70\;{\rm ms}$, the shock evolves similarly with post-bounce time $t_{\postb}$ for all simulations. In each simulation, the shock forms and propagates outward to 140 km at which point it stalls. 
After $t_{\postb} \gtrsim 70$~ms, differences in the shock evolution between the four simulations start to emerge. 
The s15 simulation has the earliest shock revival (at $\sim 80$~ms post bounce), only $\sim 10$~ms after the shock front starts to deform which can be seen in the figure as the increasing spread between the minimum and maximum shock radius. 
The s20 and s27 simulations behave similarly, albeit with shock revival and the emergence of asphericity occurring later than in the s15 simulation. 
Moreover, there is a longer delay of $\sim 60$~ms between when the shock starts to deviate from spherical symmetry and the clear shock expansion for the s20 and s27 simulations. 
At the end of our s15, s20 and s27 simulations, the shock appears to be irrevocably expanding outwards. 
Following the common definition for successful explosions found in the literature (when the shock reaches 500~km), these models could be called successfully exploding, even though it is too early for the total energy in the gain region and above to have reached positive energies by $t_{\postb} \sim 200\;{\rm ms}$. 
Finally, the s23 simulation does not show shock revival during the simulation time ($\sim 205$~ms post bounce). 
Instead, the shock radius hovers around 100~km until the end of the simulation. 
This is qualitatively similar to the initial shock evolution of a different 23~\msun model (from \citet{2018ApJ...860...93S}) which follows the non-exploding models for the first 300~ms, then continues to hover around 150~km, and only starts expanding around 600~ms after bounce.

Figure \ref{fig:entropy} presents snapshots of the fluid entropy per baryon for all four simulations. 
In each snapshot, we observe low-entropy material ahead of the shock, and also in the center of the computational domain, where the proto-neutron star is forming. 
The higher-entropy region in-between contains the gain layer, which we define as the region interior to the shock where the entropy exceeds $5\,{\rm k_{B}}/\text{baryon}$, the density is less than $\rho < 10^{10}\;{\rm g/cm^3}$, and the energy gained from electron neutrino and antineutrino absorption exceeds the energy lost by neutrino emission. 
We will focus our attention on this gain region in the next sections and going forward, we have truncated our time-series data to begin when we have qualitatively determined a meaningful gain region has formed (at $t_{\postb} \approx 61$~ms).

In Figure \ref{fig:entropy} we demonstrate the evolution of the supernovae using selected snapshots from each simulation.
The first three columns correspond to $t_{\postb} = $100, 150, 195 ms post bounce and the last column represents the final simulation time for each simulation. 
In the case of s15, the shock reached 500~km before 195 ms post bounce and thus the final panel is left blank. 
In the s15, s20 and s27 simulations, the spherical symmetry of the simulation is visibly broken when neutrino heating initiates convection in the gain region.
By 150~ms post bounce, the shocks in these three simulations are noticeably aspherical.
By $\sim$195 ms post bounce, low-entropy downflows through the gain layer amid large convective plumes are clearly visible. 

For the s23 simulation the spherical symmetry remains largely preserved for the entire duration of the simulation. 
Although some of the material in the gain region reaches comparable specific total energies as in the other three simulations, the energy gained by neutrino absorption is clearly not enough to revive the shock in this case. 
The reader may notice that the center of the quasi-circular shock front is slightly displaced from the origin of the coordinates.
Indeed, from animations of the entropy as a function of post-bounce time, we observed an oscillatory motion of the shock starting at $t_{\postb} \sim 150\;{\rm ms}$ characteristic of a SASI \citep{2003ApJ...584..971B,2007Natur.445...58B}. 
This timescale is similar to the timescale of a spiral SASI in a magnetized 24~\msun model by \citet{2025MNRAS.536..280N}.

\begin{figure}
    \centering
    \includegraphics[width=0.45\textwidth]{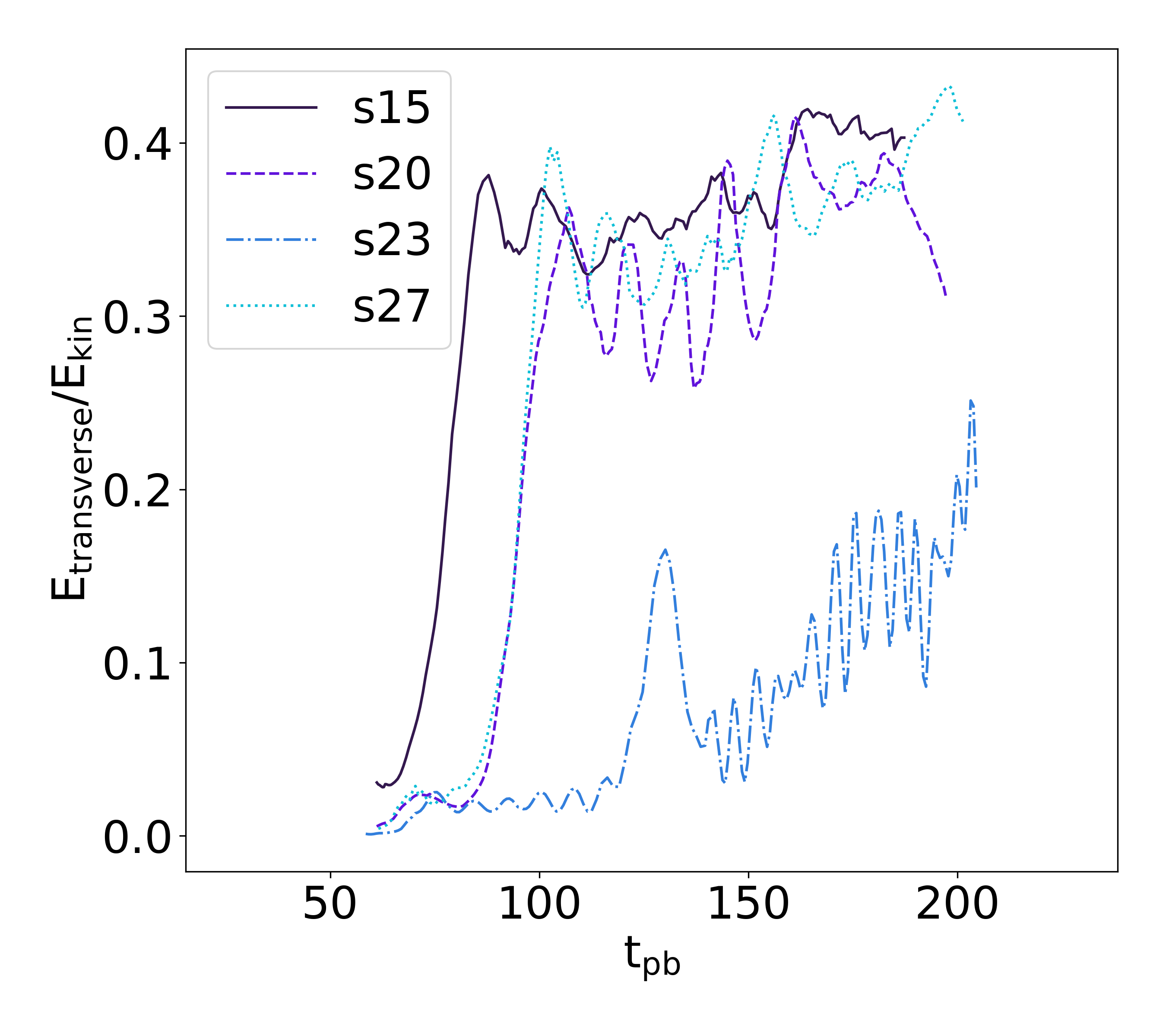}
    \caption{Ratio of the transverse energy to the total kinetic energy in the gain region as a function of the post-bounce time for the s15 (solid), s20 (dashed), s23 (dot-dashed), and s27 (dotted) simulations.    
    \label{fig:EandtransE}
    }
\end{figure}

We can be more quantitative about the time at which convection begins in each simulation by computing the fraction of kinetic energy within the gain layer which is in the non-radial directions:
\begin{widetext}
\begin{equation}
E_{\mathrm{transverse}} \,/\, E_{\mathrm{kin}}  
= \int_{V_{\gain}} \rho\,\left( v^2_{\theta} + v^2_{\phi} \right) \,dV \Biggm/ \int_{V_{\gain}} \rho\,\left( v^2_{r} + v^2_{\theta} + v^2_{\phi} \right) \,dV,
\end{equation}
\end{widetext}
where $v_r,\ v_{\theta},\ v_{\phi}$ are the radial, polar and azimuthal components of the velocity and $V_{\gain}$ is the volume of the gain region, defined above.
This ratio is shown in Figure \ref{fig:EandtransE}.
For the s15, s20, and s27 simulations, the fraction of the kinetic energy in the transverse (i.e., non-radial) directions rises very rapidly between $t_{\postb} \sim 60\;{\rm ms}$ and $t_{\postb} \sim 100\;{\rm ms}$ as the shock stalls. 
The rise is earlier in the s15 simulation than in the s20 and s27 simulation, but by $t_{\postb} \sim 100\;{\rm ms}$ they all reach the level of 30-40\% with a slow growth of the ratio thereafter. 
Interestingly, the figure shows that the fraction of kinetic energy in the transverse directions begins to grow in the s23 simulation starting at $t_{\postb} \sim 120\;{\rm ms}$. 
Even though the s23 simulation is not exploding --- the shock is not seen to be advancing in Figure \ref{fig:shockradii} and we shall shortly show more evidence that further indicates this conclusion --- the high densities and large angular motions present in this simulation due to the SASI means the ratio of the transverse and total kinetic energy in the gain region becomes comparable to the same ratio of energies in the successful explosions.  
While it does not reach the $\sim 40$\% level seen in the other three simulations, it is only smaller by a factor of $\sim 2$ at $t_{pb} = 200\;{\rm ms}$.

\begin{figure}
\centering
\includegraphics[width=0.45\textwidth]
{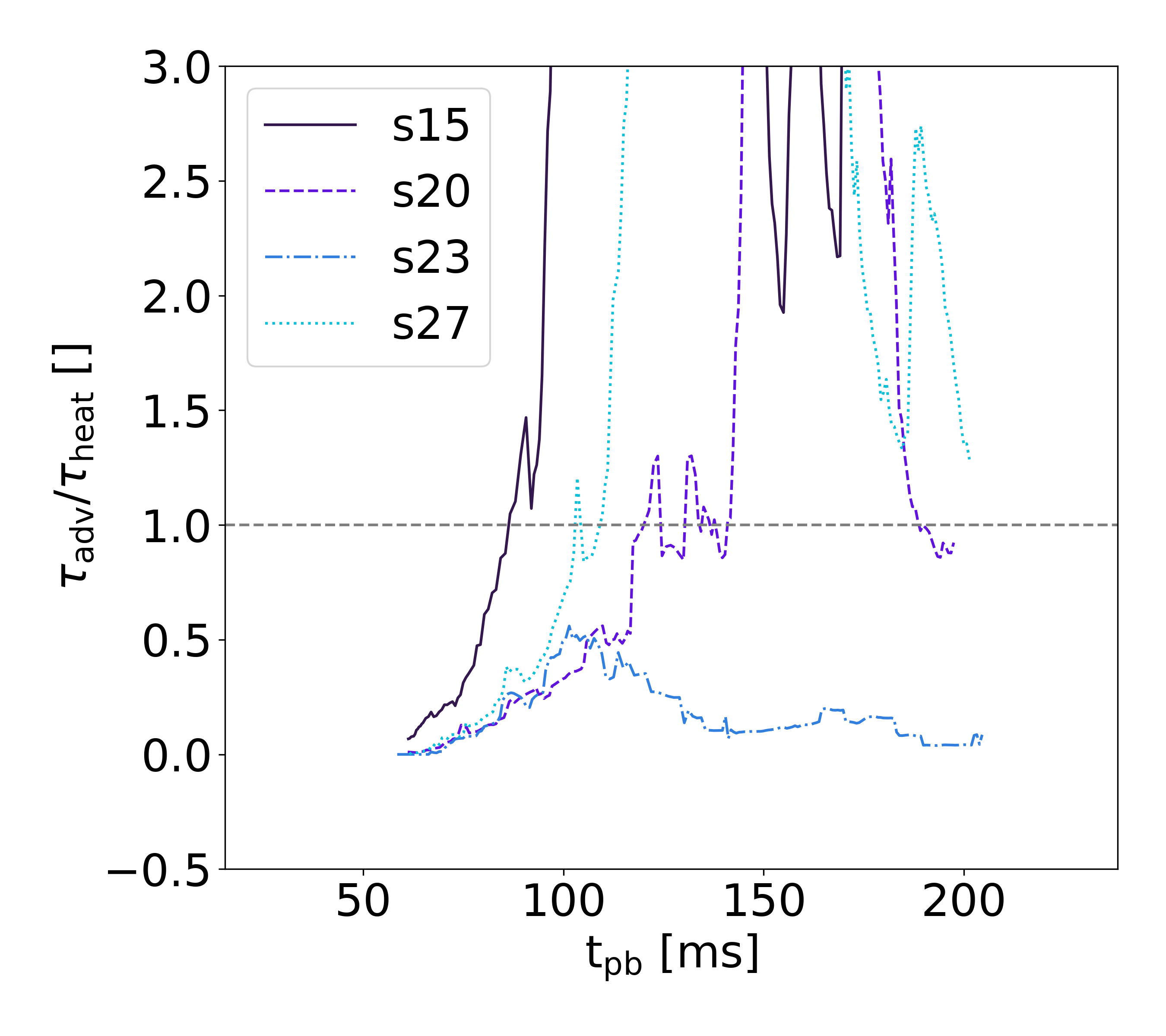}
\caption{
The ratio of advection timescale to heating timescale as a function of the post-bounce time for all four simulations. The gray dotted line marks where the ratio equals unity.
\label{fig:ratio}
}
\end{figure}

We want to turn our analysis to the mechanism by which the shock is revived in our simulations. 
In a neutrino-driven explosion, the shock is revived due to neutrino-heated material in the gain region being brought up to the shock by convection and buoyancy. 
Shock revival occurs when the material within the gain region has sufficient time to absorb enough of this heating energy to overcome gravity.
Thus the revival time can be estimated by comparing the heating timescale to the advection timescale through the gain region. 
For the advection timescale we adopt the definition used in \citet{2006A&A...457..281B}: 
\begin{equation}
    \tau^*_{\mathrm{adv}}(t_1) = t_2(M) -t_1(M),
    \label{eq:advt}
\end{equation}
where $t_2(M)$ is the time at which a mass $M$ is enclosed by the shock and $t_1(M)$ is the time at which that same mass falls beneath our gain region.
The neutrino heating timescale is defined as the amount of time that material in the gain region would need to be exposed to the current neutrino heating rate to become gravitationally unbound, i.e.,
\begin{equation}
    \label{eq:tnu}
    \tau_{\mathrm{heat}} = \frac{|E_{\gain}|}{\dot{Q}_{\mathrm{heat}}},
\end{equation}
where $E_{\gain} = E_{\mathrm{kin}}+E_{\mathrm{int}}+E_{\mathrm{grav}}$ is the total energy in the gain region, $E_{\mathrm{int}}$ and $E_{\mathrm{grav}}$ are, respectively, the  total internal and total gravitational energy of the fluid in the gain region, and $\dot{Q}_{\mathrm{heat}}$ is the total neutrino heating rate in the gain region.
It is found 
that when the ratio of the advection timescale to the heating timescale, $\tau_{\mathrm{adv}}/\tau_{\mathrm{heat}}$, exceeds unity, the stalled shock is considered revived and, on average, the shock begins to move outwards due to the increased energy of the enclosed material \citep{1998suco.conf....7J,2001A&A...368..527J,2005ApJ...620..861T,2008ApJ...688.1159M}. 
Note that this does not guarantee an explosion, but simply is an indicator of an outward moving shock.
Figure \ref{fig:ratio} shows the ratio of these two timescales for all four simulations, with the line for a ratio of 1 highlighted in gray 
From Figure \ref{fig:ratio}, we see that this ratio exceeds unity in the s15, s20 and s27 simulations at times which closely match the shock revival times seen in Figure \ref{fig:shockradii}, supporting a neutrino-driven explosion mechanism for these simulations. 
For the s23 simulation, the ratio never exceeds unity, which is consistent with the lack of a shock revival in this simulation.

\begin{figure}
\centering
    \includegraphics[width=0.5\textwidth]{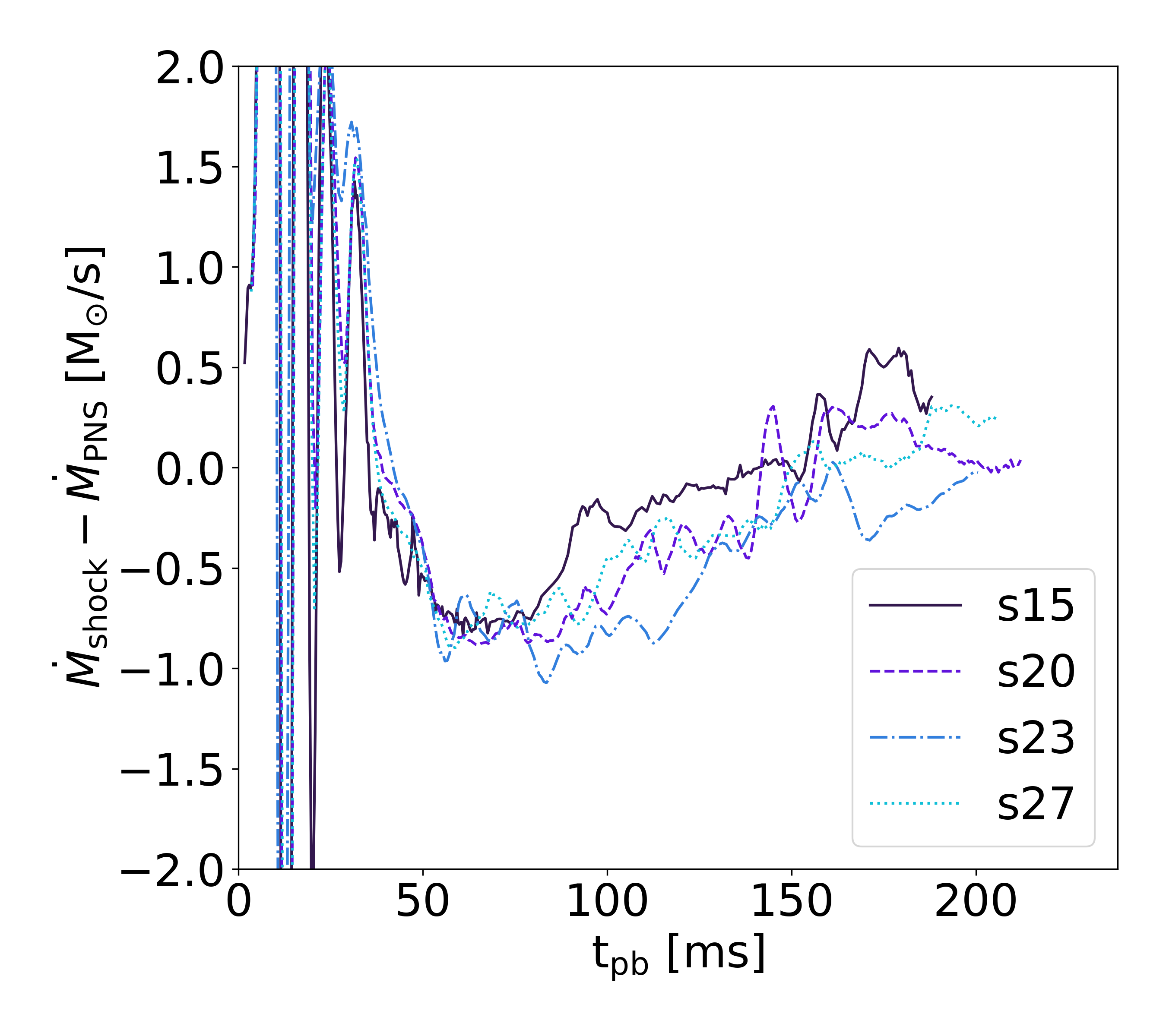}   
\caption{
Difference between the mass accretion rate through the shock front ($\dot{M}_{\mathrm{shock}}$) and the mass accretion rate onto the PNS ($\dot{M}_{\mathrm{PNS}}$) as a function of the post-bounce time for all four simulations. The data is smoothed over a 5 ms window. 
\label{fig:massaccretion}
}
\end{figure}

Another diagnostic invoked to investigate the phase leading up to shock revival consists of examining the mass inflow rate through the shock front ($\dot{M}_{\mathrm{shock}}$) and the mass inflow rate onto the PNS ($\dot{M}_{\mathrm{PNS}}$). 
The comparison of these rates are shown in Figure \ref{fig:massaccretion}.
Initially, $\dot{M}_{\mathrm{shock}}$ and $\dot{M}_{\mathrm{PNS}}$ are firmly related. 
In the first few tenths of milliseconds after bounce, $\dot{M}_{\mathrm{shock}}$ decreases more rapidly than $\dot{M}_{\mathrm{PNS}}$ and the initial shock turns into a standing accretion shock. 
Once non-radial fluid motion starts to develop around 60--100~ms after bounce, we observe a drop in $\dot{M}_{\mathrm{PNS}}$ (while $\dot{M}_{\mathrm{shock}}$ continues to decrease smoothly), resulting in the increase of $\dot{M}_{\mathrm{shock}} - \dot{M}_{\mathrm{PNS}}$. 
Once the stalled shock starts expanding again, $\dot{M}_{\mathrm{PNS}}$ is smaller than $\dot{M}_{\mathrm{shock}}$ and $\dot{M}_{\mathrm{shock}} - \dot{M}_{\mathrm{PNS}}$ reaches zero or slightly positive values. 
While the difference $\dot{M}_{\mathrm{shock}} - \dot{M}_{\mathrm{PNS}}$ is similar for all four simulations, the individual amount of accretion through the shock ($\dot{M}_{\mathrm{shock}} - \dot{M}$) and onto the PNS ($\dot{M}_{\mathrm{PNS}}$) are higher for the s23 model (no shock expansion) than for any of the other models (with shock expansion), by up to a factor of 2 after $\sim 90$~ms post bounce

In summary, the results from our simulations shown in Figures \ref{fig:shockradii} through \ref{fig:massaccretion} are broadly consistent with other 3D simulations
\citep{
2013ApJ...778L...7C,
2014ApJ...785..123C,
2014PhRvD..90d5032T,
2015arXiv150705680M,
2017MNRAS.468.2032A,
2018MNRAS.477.3091V,
2019ApJ...873...45G,
2020MNRAS.498L.109M,
2023PhRvD.107j3015V,
2024MNRAS.531.3732S,
2024MNRAS.532.4326P,
2025arXiv250411537S}.


\section{Measuring the turbulence} \label{sec:turbKE}

The turbulence in the gain region is conjectured to be an important ingredient in revival of the shock \citep{2015ApJ...799....5C}. 
The prevailing theory being that the average adiabatic index of the turbulence (which we shall define later in Section \ref{sec:TurbulentAdiabticIndex}) is larger than in a non-turbulent fluid \citep{2015ApJ...801L..24M,2015ApJ...799....5C,radice_turbulence_2018,2021Natur.589...29B,2024MNRAS.528L..96M}. 
To investigate how the turbulence affects the simulations, it is first necessary to define a measure of the amount of turbulence. 
This is not a straight-forward task; as discussed in \citet{2000ExFl...29..275A}, there is no one, perfect measure of turbulence. 

In some previous studies of turbulence in supernovae \citep{2012ApJ...755..138H,2013ApJ...765..110D,2014MNRAS.440.2763F,2014ApJ...785..123C} the reader will find the turbulent kinetic energy $E_{\turb}$ defined to be equivalent to the non-radial kinetic energy, i.e.,
\begin{equation}
E_{\turb} \equiv E_{\mathrm{transverse}} = \int_{V_{\gain}} \frac{\rho}{2}\,\left( v^2_{\theta} + v^2_{\phi} \right) \,dV. 
\end{equation}
This definition of the turbulent kinetic energy obviously does not include any contribution from flow in the radial direction which could be turbulent, but it does include a contribution from the non-radial flow which may not be turbulent; e.g., flows that are simply the lateral motion of the fluid at the top of convective plumes and/or any global rotational motion of the fluid due to a SASI.
We shall not consider this definition further in this paper. 

Another commonly encountered approach for measuring the amount of energy stored in turbulence is to invoke a Reynolds decomposition of the velocity field and then define the turbulent kinetic energy as  
\begin{equation}
E_{\turb} = \int_{V_{\gain}} \frac{\rho}{2}\,|\boldsymbol{v} - \langle \boldsymbol{v} \rangle |^2\,dV,
\end{equation}
where $\langle \boldsymbol{v} \rangle$ is an expectation value for the velocity.
Note that $\langle \boldsymbol{v} \rangle$ can be a function of position and time.
Defining the expectation value is not straight-forward; in statistically steady-state flows, the expectation value for the velocity $\langle \boldsymbol{v} \rangle$ is the time-averaged velocity of the fluid at a given location.
While it is certainly possible to time-average fluid properties in a supernova simulation, it is conceptually difficult to accept them as useful because the system is not close to a statistical steady-state (even during the period when the shock has stalled). 
Instead, authors usually substitute a spatial average of the the fluid velocity in place of a time average at a given location. 
However, spatial averages are just as conceptually fraught in situations where the fluid may have organized large-scale motions, i.e., on scales close to the dimensions of the fluid volume. 
In the case of a supernova, these large scale motions are due to convection and/or SASI. 
Since there is no clear way to define $\langle \boldsymbol{v} \rangle$, we have considered two different spatial average options. 

The first method we adopt we denote as the Spherical Spatial Average (SSA) and is based on \citet{2015ApJ...799....5C}.
The SSA algorithm for computing $\langle \boldsymbol{v} \rangle$ is based on the observation that the general flow of the fluid in the gain region is toward the PNS.
Thus, we define $\langle \boldsymbol{v} \rangle$ as $\langle v_r \rangle\,{\bf {\hat r}}$ where $\langle v_r \rangle$ is the average radial velocity for all zones at a given radius $r$. 
Since we do not use a spherical coordinate grid in our simulations, we implement this definition by finding all the Cartesian grid cells with a cell-center radius in a given 1 km thick shell.
Thus, we define 
\begin{equation}
\langle v_r \rangle = \frac{1}{M_{\mathrm{shell}}}\,\int_{V_{\mathrm{shell}}} \rho\,v_r \,dV ,
\end{equation}
with 
\begin{equation}
M_{\mathrm{shell}} = \int_{V_{\mathrm{shell}}} \rho\,dV.
\end{equation}
Note that when the shock becomes aspherical, only grid cells in the gain region are used to compute the average radial velocity, thus the shell no longer forms a complete spherical surface.

The second method we consider we call the Local Spatial Average (LSA) method. 
In this method, we define the expectation value of the velocity, $\langle \boldsymbol{v} \rangle$, for a particular grid cell as being the mass-weighted average of the velocity in a cube surrounding the cell:
\begin{equation}
\langle \boldsymbol{v} \rangle = \frac{1}{M_{\mathrm{cube}}}\,\int_{V_{\mathrm{cube}}} \rho\,\boldsymbol{v} \,dV ,
\end{equation}
where 
\begin{equation}
M_{\mathrm{cube}} = \int_{V_{\mathrm{cube}}} \rho\,dV
\end{equation}
and $V_{\mathrm{cube}}$ is a cubic volume of side length $L$ centered on the grid cell under consideration. 
Note that when we perform the integral, the volume ${V_{\mathrm{cube}}}$ only includes those cells which are found to be in the gain region, implying that the volume is not always a complete cube. 
The size of the averaging volume must be chosen appropriately.
In the limit where the size of the averaging volume becomes small, $\langle \boldsymbol{v} \rangle$ approaches the velocity of the grid cell under consideration, in which case the deviation of the velocity from the expectation value approaches zero, i.e., $|\boldsymbol{v} - \langle \boldsymbol{v} \rangle| \rightarrow 0$ and therefore the turbulent kinetic energy also approaches zero in this limit. 
We must select a suitable length scale for which the turbulent kinetic energy is captured. 
\begin{figure*}
     \centering
     \includegraphics[width=0.48\textwidth]{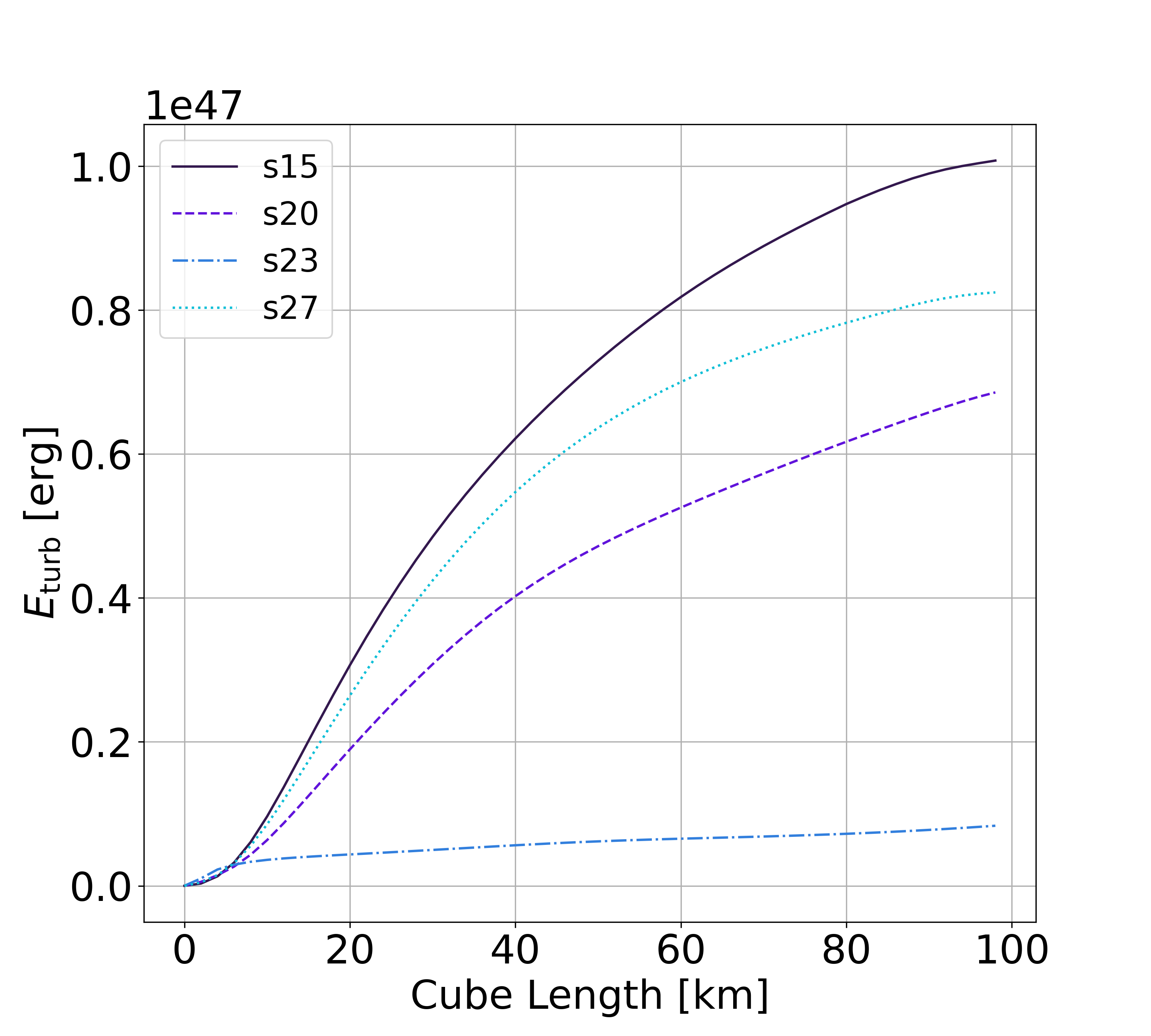}
     \includegraphics[width=0.48\textwidth]{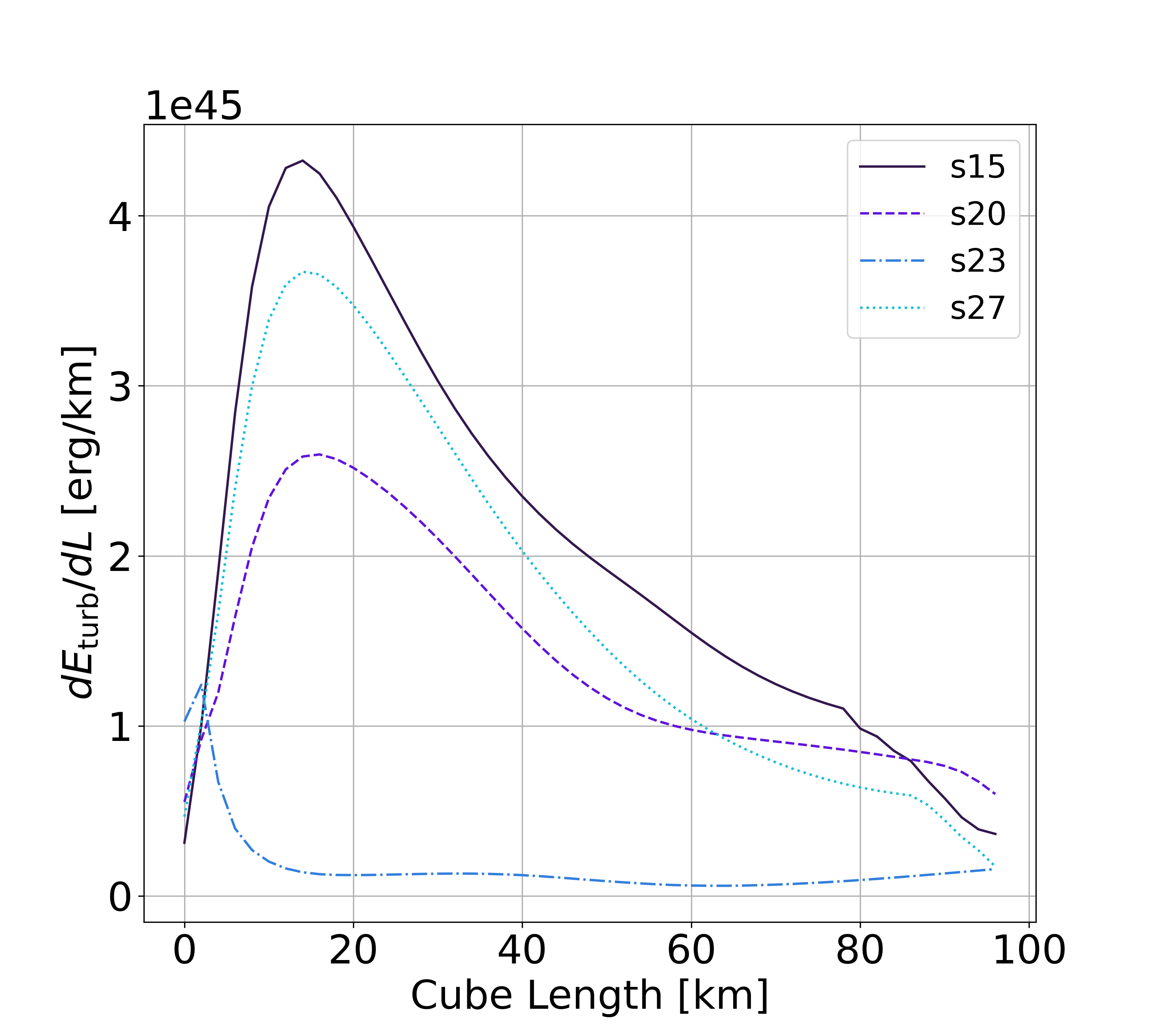}
    \caption{
    The total turbulent kinetic energy (left) and the corresponding slope of the total kinetic energy (right) as a function of the cube side length used in the LSA method at $t_{\postb} \approx 150$ ms for all four simulations.
    \label{fig:cube_slope}
    }
\end{figure*}
The turbulent kinetic energy in the gain region for a snapshot at $t_{\postb} \approx 150\;{\rm ms}$ computed using the LSA method as a function of the averaging volume scale $L$ is shown in the left panel of Figure \ref{fig:cube_slope}.
As described above, when the averaging volume is small, the turbulent kinetic energy approaches zero. 
As the averaging volume grows we see a rise in the turbulent kinetic energy but we do not observe a plateau, implying there is no range of averaging volume sizes where the turbulent kinetic energy is independent of the size of the averaging volume. 
At $L \sim 20 \;{\rm km}$ we find a maximum in the derivative of $E_{\turb}$, seen as peak in right panel of Figure \ref{fig:cube_slope}. 
Although there is not a clear range over which the turbulent kinetic energy is independent of the length scale, for the purposes of defining the turbulent kinetic energy using the LSA method we have adopted a cube length scale of $20\;{\rm km}$. 
We shall later show that this averaging volume scale is larger than the characteristic vortex radius (see Section \ref{sec:enstrophycomparison}).

Finally we adopt a definition of the turbulent kinetic energy we refer to as the Spectral method based upon the approach found in \citet{2013PhST..155a4022E}.
A variation of this method can be found in \citet{2020PhyS...95f4005C}.
The Spectral Method forgoes trying to define $\langle \boldsymbol{v} \rangle$ directly and instead removes the large-scale fluid motion by excluding the small wavenumbers from a Fourier transform and then integrating over the remaining range of wavenumbers. More precisely, let $\hat{\boldsymbol{f}}$ be the Fourier transform of an arbitrary vector quantity $\boldsymbol{f}$ in the gain region, 
\begin{equation}
\hat{\boldsymbol{f}} = \frac{1}{(2\pi)^3}\,\int_{V_{\gain}} \boldsymbol{f}\,\exp\left(2\imath \pi\,\boldsymbol{k}\boldsymbol{\cdot}\boldsymbol{x}\right)\,d^{3}x.
\label{eq:fourier}
\end{equation}
Integrating the squared magnitude of the transformed quantity over a spherical shell in the $k$-space, one obtains the power spectrum
\begin{equation}
\hat{e}_f(k) = \frac{1}{2} \int_{\mathrm{k-shell}} |\hat{\boldsymbol{f}}|^2 k^2\,d\Omega_k.
\label{eq:kshell}
\end{equation}
Adopting $\boldsymbol{f} = \sqrt{\rho}\,{\boldsymbol{v}}$ and including only data originating from the gain layer gives the power spectrum $\hat{e}_{ \sqrt{\rho}\,{\boldsymbol{v}}}$ of the kinetic energy in the gain region. The \emph{turbulent} kinetic energy is defined as the integral of the power spectrum above a minimum wavenumber, i.e.
\begin{equation}
E_{\turb} = \int_{k_{\mathrm{min}}}^{\infty} \hat{e}_{ \sqrt{\rho}\,{\boldsymbol{v}}} \,dk .
\label{eq:kbound}
\end{equation}
Here, $k_{\mathrm{min}}$ is the minimum wavenumber considered. 
As with the LSA method described above, we face the question of what to use for the minimum wavenumber $k_{\mathrm{min}}$.
\citet{2013PhST..155a4022E} adopted $k_{\mathrm{min}} = 0.1 \boldsymbol{\times} (2\pi) \,{\rm km}^{-1}$; however, we have found that for our simulations this lower bound does not sufficiently capture the onset of convection, which was not present in \citet{2013PhST..155a4022E}.
As such, we have followed their same method to determine a $k_{\mathrm{min}}$ suitable for our simulations.

\begin{figure}
\centering
\includegraphics[width=0.5\textwidth]
{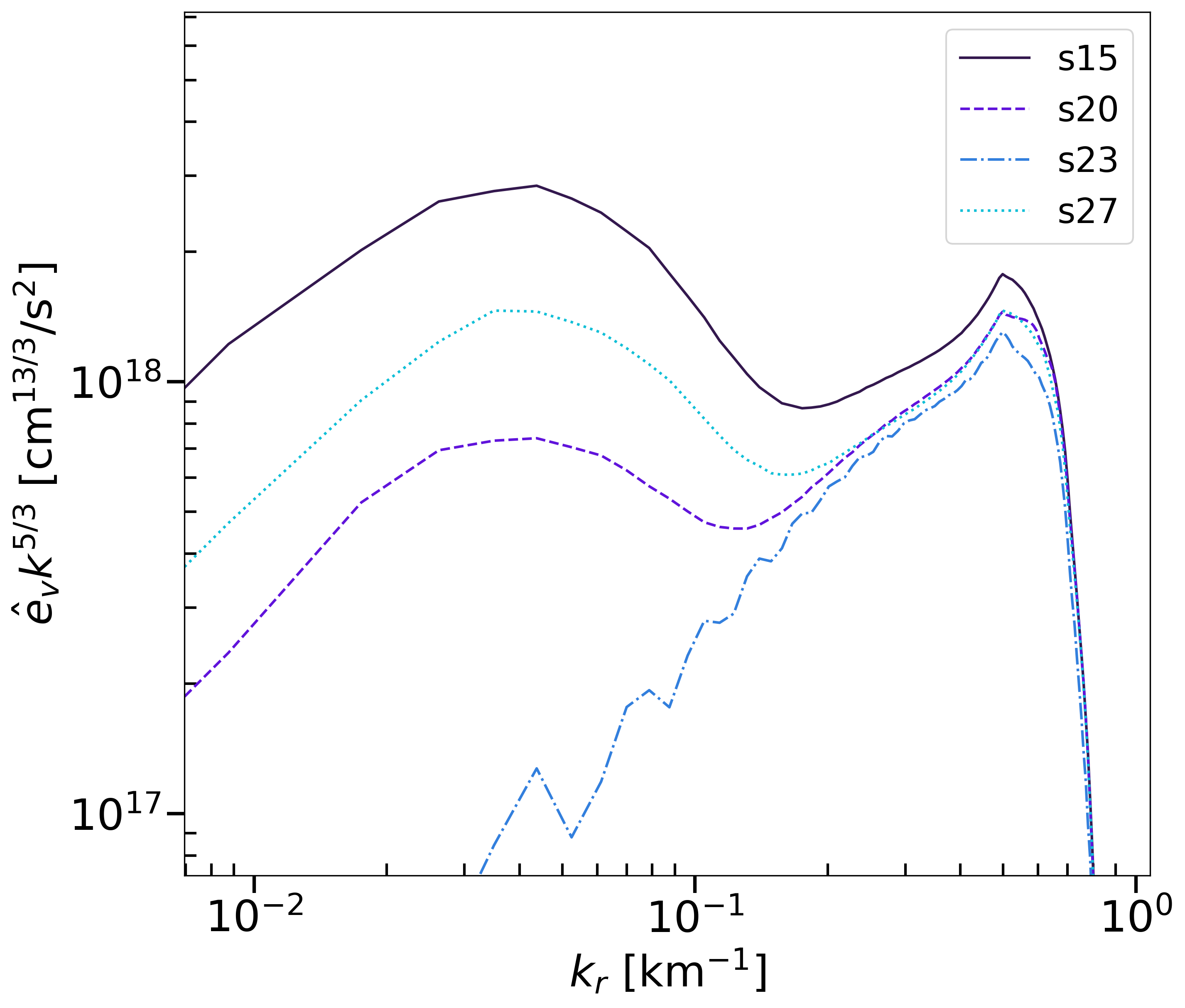}
\caption{
The compensated specific kinetic energy power spectrum at $t_{\postb} = 150$ ms for all four simulations. 
\label{fig:powerspec}
}
\end{figure}

If the fluid has well-developed turbulence, we expect the compensated spectral specific kinetic energy, defined to be $\hat{e}_{v}\,k^{5/3}$ where $\hat{e}_{v}$ is the spectrum of the specific kinetic energy, to be a constant for a range of wave numbers $k_r$. 
We have plotted this compensated specific kinetic energy power spectrum in Figure \ref{fig:powerspec} for all four simulations at $t_{\postb} \approx 150$ ms. 
This figure can be compared with the left panel of Figure 3 in \citet{2013PhST..155a4022E}, although we caution the reader that a) the definition of $k$ in \citet{2013PhST..155a4022E} means the wavenumbers there are $2\,\pi$ larger than those used in this paper, and b) we show a range of wavenumbers shifted to slightly higher values. 
Examining Figure \ref{fig:powerspec} we do not see a horizontal section - a region of wavenumbers where the compensated spectral specific kinetic energy is constant -  in any of the curves in our plot, but in the three successful explosions we find: a minimum in the spectrum at $k_r = 0.1 - 0.2 \;{\rm km}$ (corresponding to a physical scale of 5 - 10 km), and a maximum at $k_r \sim 0.04;{\rm km}$ (corresponding to a physical scale of 25 km). For the purposes of defining the turbulent kinetic energy using the Spectral method we set the lower bound of integration $k_{\mathrm{min}}$ to be $k_{\mathrm{min}} = 0.025\,{\rm km}^{-1}$, equivalent to a physical scale of 40 km.

\section{Application to Simulations}
\label{sec:results}

\subsection{Turbulent Kinetic Energy}

\begin{figure*}
\centering
\includegraphics[width=\textwidth]{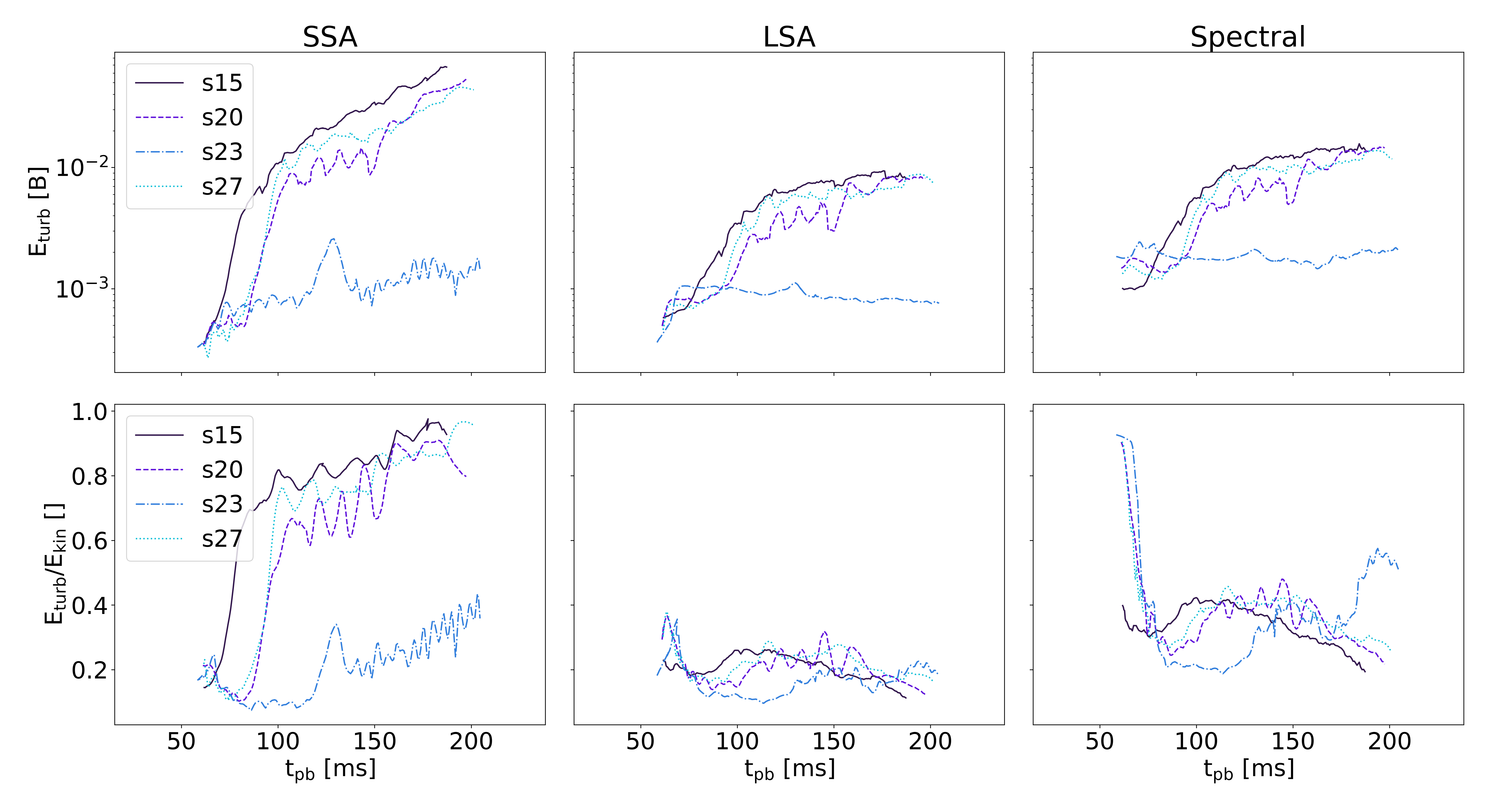}
\caption{Top: The turbulent kinetic energy in the gain region as a function of the post-bounce time for all four simulations. 
Bottom: Ratio of the total turbulent kinetic energy to the total kinetic energy in the gain region as a function of the post-bounce time for all four simulations. 
Each column represents a different method for computing the turbulent kinetic energy. 
Left: Spherical Spatial Average (SSA) method; 
Middle: Local Spatial Average (LSA) method; 
Right: Spectral Method. 
\label{fig:TurbE}
}
\end{figure*}

We compute the total turbulent kinetic energy in the gain region using the three methods described above. The results are shown in the top row of Figure \ref{fig:TurbE}. 
No matter the method used, for a successful explosion, there is a rapid increase in the amount of turbulent kinetic energy in the gain region when convection begins (80--100 ms after bounce). 
Thereafter, the increase is more gradual in the interval $100\;{\rm ms} \lesssim t_{\postb} \lesssim 200\;{\rm ms}$.
While the general trend of the results is the same across all methods, we note that the LSA and Spectral methods give values consistently smaller than the SSA method with the consequence that 
the growth in the turbulent kinetic energy during the interval $100\;{\rm ms} \lesssim t_{\postb} \lesssim 200\;{\rm ms}$ using the LSA and Spectral methods is only by a factor $\sim 2$ compared to a factor closer to $\sim 8$ over the same period when using the SSA method. 

The s23 simulation is clearly distinct from the successful explosions across all three methods. 
The amount of turbulent kinetic energy is smaller for s23 by approximately an order of magnitude (or more) compared to s15, s20 and s27. 
Moreover, there is little (if any) growth of the turbulent kinetic energy over the period $100\;{\rm ms} \lesssim t_{\postb} \lesssim 200\;{\rm ms}$ for the s23 simulation.

The difference between the amount of turbulent kinetic energy computed in the s23 simulation compared to the s15, s20, and s27 simulations is largely due to the difference in size / mass in the gain regions of the simulations.
To eliminate this dependence, we show, in the bottom row of Figure \ref{fig:TurbE}, the evolution of the ratio of the turbulent kinetic energy to the total kinetic energy within the gain region, using each of the three methods.
These panels reveal there are fundamental differences between the methods already hinted at by the previously noted differences in the growth rates of the turbulent kinetic energy in the top panels. 
Using the SSA method there is still a clear distinction between a successful and a failing supernova in the energy ratio, but this distinction disappears in the ratios computed using the LSA and Spectral methods. 
Note that the ratio of turbulent kinetic energy to total kinetic energy using the SSA method has strong resemblance to Figure \ref{fig:EandtransE}.
When the SSA method is used, the ratio of turbulent kinetic energy to total kinetic energy is of order $\sim 60\%$ in successful explosions at the onset of the explosion and later rises to $\sim 80\%$ by the end of the simulation. The s23 stands apart from the other runs.
Although the ratio of kinetic energies in s23 using the SSA method does not reach the same level as the successful explosions, there is still an overall growth of the ratio after $t_{\postb} \sim 120\;{\rm ms}$ to $\sim 40\%$ by the end of the simulation.
In contrast, the ratio of turbulent to total kinetic energy computed using the LSA or Spectral method are very similar for all \emph{four} simulations.
There is no substantial distinction between successful and failing simulations. 
Aside from some temporary fluctuations, the ratio of energies lies between $\sim 20\%$ and $\sim 40\%$ for all four simulations after $t_{\postb} \sim 80\;{\rm ms}$, with no apparent change during shock revival of the s15, s20 and s27 simulations. 
One feature of note is that at approximately 195 ms post bounce, there is an uptick in the percentage of turbulent energy calculated by the spectral method for s23.
We posit that this uptick is related to the SASI that occurs in the s23 at late times.
As the large scale angular motion of the shock increases, these SASI modes begin to be detected by the Fourier Transform. 

The different methods for computing the turbulent kinetic energy clearly do not capture the same physics, and the differences would lead to different conclusions about the relative importance of turbulence for successful explosions. 
To further probe this, we turn our attention to other fluid quantities which are also associated with turbulence. 
For this, we focus on the 3 simulations that have a substantial amount of turbulence in the gain region; the s15, s20, and s27 simulations.


\subsection{Enstrophy Comparison} 
\label{sec:enstrophycomparison}
\begin{figure*}
 \centering
\includegraphics[width=\textwidth]{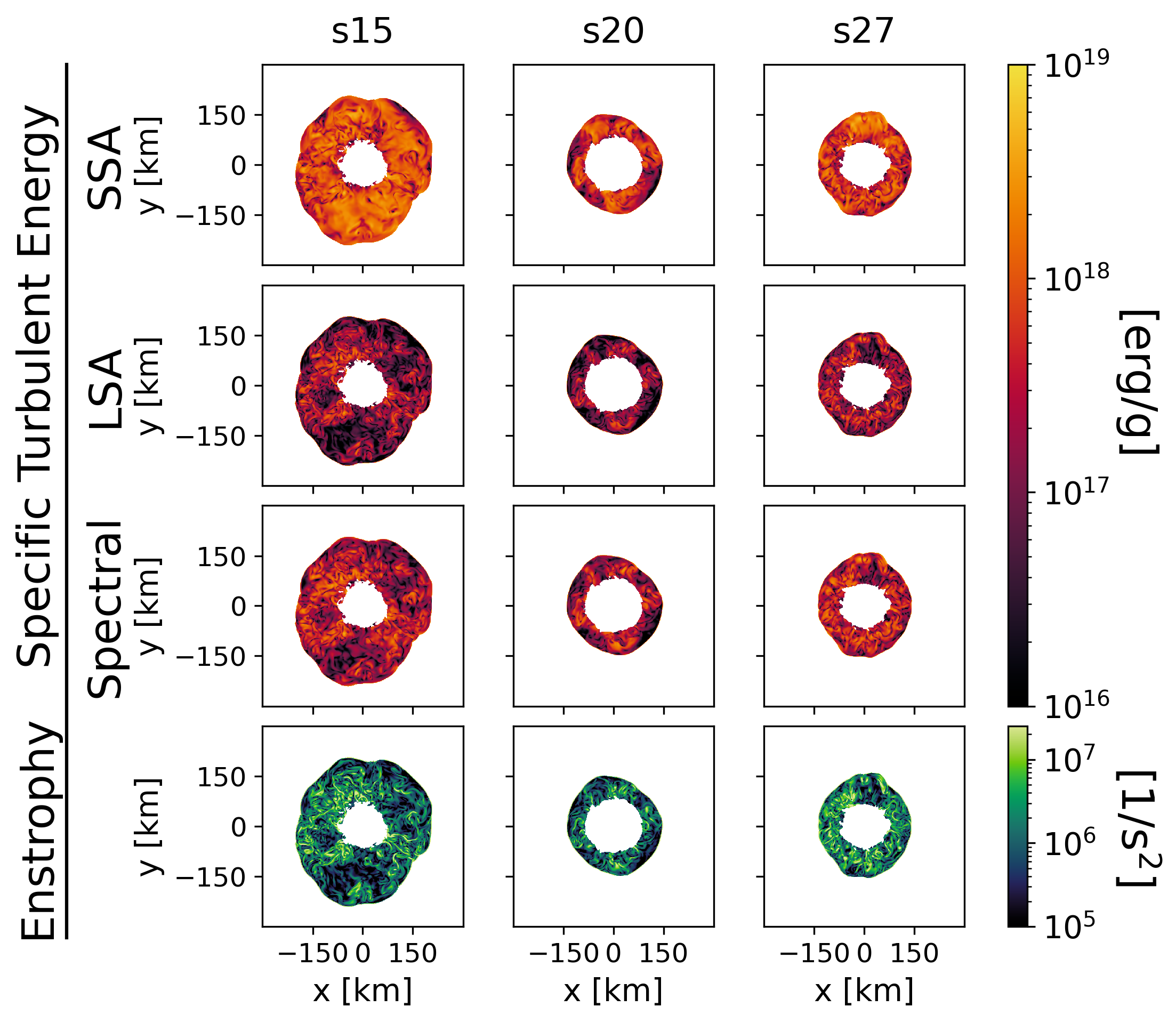}
\caption{
Spatial maps of the specific turbulent energy in the $xy$-plane computed using all three methods (First row: SSA; Second row: LSA; Third row: Spectral Method) for the exploding simulations. 
Bottom: Spatial maps of the enstrophy in the $xy$-plane for the same simulations.
All panels are at 150~ms post bounce. 
\label{fig:specific_eturb}
}
\end{figure*}

Fluid vorticity, $\boldsymbol{\omega} = \boldsymbol{\nabla} \boldsymbol{\times} \boldsymbol{v}$, or alternatively enstrophy $\epsilon =|\boldsymbol{\omega}|^2 /2$, are also regarded as intrinsic aspects of turbulence \citep{2004iit..book.....T} and thus a good, but not sufficient, measure of its presence. 
Therefore, one might expect a map of the enstrophy should resemble a map of the specific turbulent kinetic energy $e_{\turb}$.
We present the spatial maps of the specific turbulent kinetic energy at the snapshot time of $t_{\postb} = 150\;{\rm ms}$ in Figure \ref{fig:specific_eturb}. 
These spatial plots are masked to show only the values in the gain region.
For the Spectral method we generate the map by Fourier transforming the velocity $\boldsymbol{v}$, setting the amplitude of all Fourier modes to zero for all wavenumbers $k \leq k_{\mathrm{min}}$, then transforming back to real space to produce the velocity field $\boldsymbol{u}$. 
Using $\boldsymbol{u}$, the specific turbulent kinetic energy for the Spectral method is defined as $e_{\turb} = |\boldsymbol{u}|^2 /2$. 
Finally, the map of the enstrophy is also shown in the same figure in the bottom row for direct comparison.

Comparing these figures, we observe that, using either the LSA or Spectral methods, the areas of greatest specific turbulent energy are coincident with the areas of greatest enstrophy. And similarly, areas where we find the lowest amounts of specific turbulent kinetic energy are also areas with low enstrophy. 
In contrast, the SSA method yields maps where we observe regions with large amounts of specific turbulent kinetic energy but where the enstrophy is low. These are regions where the fluid is moving outwards, such as in convective plumes. At the other end of the scale, in places where the SSA finds only a small amount of specific turbulent kinetic energy, there is also a small amount of enstrophy. Thus the amount of correlation between the maps of specific turbulent kinetic energy and enstrophy is larger when using the LSA or Spectral methods than when using the SSA. 

That the specific turbulent kinetic energy using the SSA is dominated by the convective plumes can be explained by examining how the expectation value for the velocity is defined for this method. In the SSA method, the average radial velocity $\langle v_r \rangle$ is computed from all the fluid on a spherical shell.
Given that the general flow of the fluid in the gain region is toward the PNS, $\langle v_r \rangle$ will be negative. 
The specific turbulent kinetic energy for the SSA method is $e_{\turb} = |\boldsymbol{v} - \langle v_r \rangle\,\hat{\bf{r}} |^2/2$. 
For a fluid element which is moving outwards, i.e., $\boldsymbol{v} \approx |\boldsymbol{v}|\,\hat{\bf{r}}$, the contribution to the specific turbulent kinetic energy is $e_{\turb} = (|\boldsymbol{v} - \langle v_r \rangle\,\hat{\bf{r}} |)^2/2 \approx ( |\boldsymbol{v}|^2 + \langle v_r \rangle^2 + 2\,|\boldsymbol{v}|\,|\langle v_r \rangle| )/2$.
For a fluid element with an inward velocity, i.e., $\boldsymbol{v} \approx -|\boldsymbol{v}|\,\hat{\bf{r}}$, the contribution to the specific turbulent kinetic energy is approximately $e_{\turb} = ( |\boldsymbol{v}|^2 + \langle v_r \rangle^2 - 2\,|\boldsymbol{v}|\,|\langle v_r \rangle| )/2$.
Thus, outflows dominate the specific turbulent kinetic energy in the SSA method.
In comparison, in the LSA method the expectation value for the velocity is computed from the local surrounding fluid. 
In a convective plume the expectation value will point in approximately the same direction as the velocity of the fluid element under consideration, and so the deviation of the fluid velocity in the outflow will be smaller, thus leading to a much smaller contribution to the specific turbulent kinetic energy from a convective plume.

\begin{figure*}
\centering
\includegraphics[width=\textwidth]{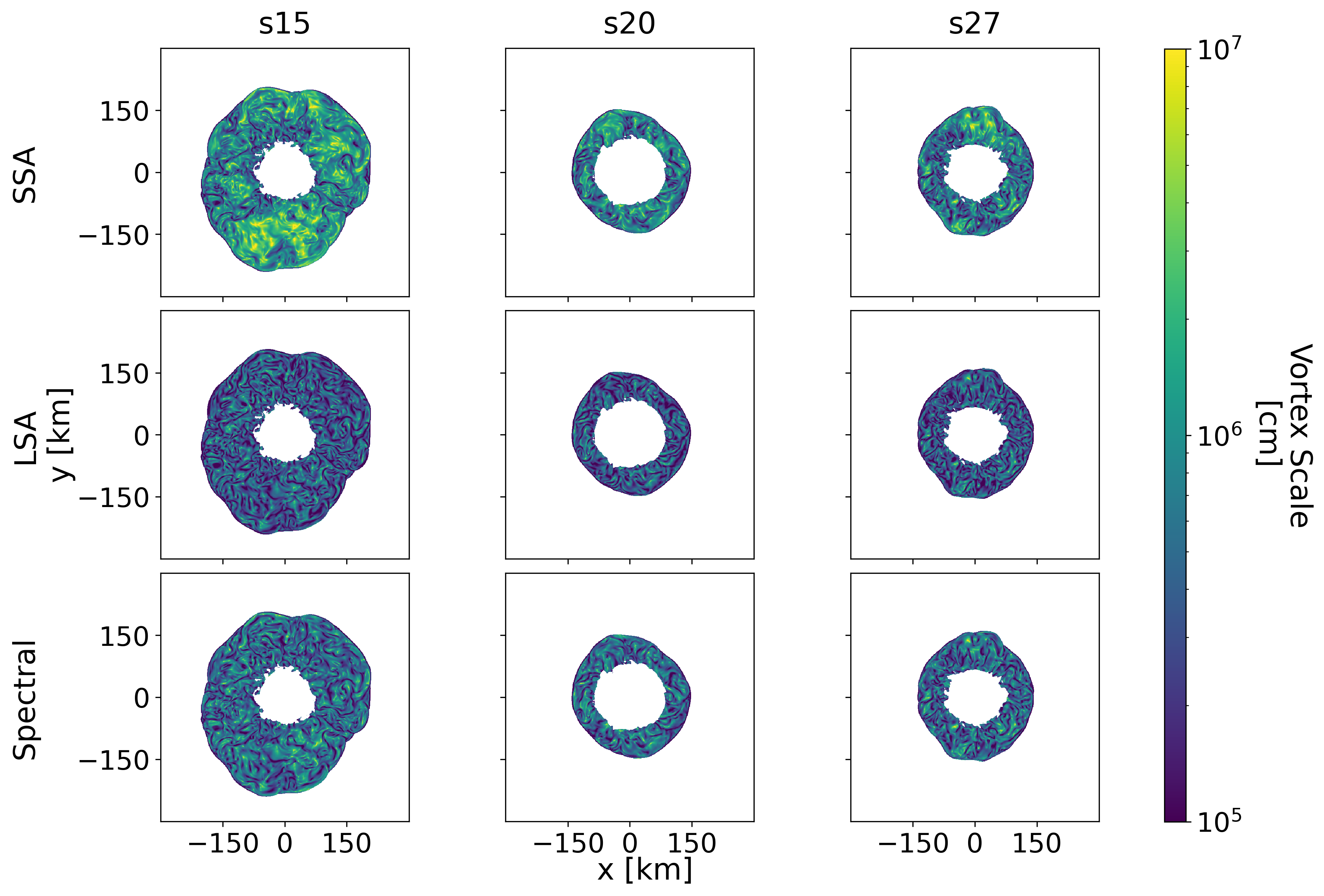}
\caption{The vortex radius scale using the SSA (top row), LSA (middle row), and spectral method (bottom row) for the s15, s20, and s27 simulations (from left to right). 
All panels are at 150~ms post bounce. 
\label{fig:vortexscale}
}
\end{figure*}
The calculation of the enstrophy also allows us to determine the characteristic scale for the turbulent vortices. 
Knowing this scale allows us to validate the choice of LSA averaging volume and the $k_{\mathrm{min}}$ chosen for the Spectral method.
We follow the method outlined in \citet{2000ApJ...545..475R} and define the characteristic vortex scale as
\begin{equation}
\ell = \sqrt{ e_{\turb} / \epsilon},
\end{equation}
where $\epsilon$ is the enstrophy. 
Note that \citet{2000ApJ...545..475R} and \citet{Endeve_2012} call this same quantity the `flow Taylor microscale'.
In Figure \ref{fig:vortexscale} we plot 2D slices of the characteristic vortex scale using all the definitions of turbulence for the same snapshots as those shown in Figure \ref{fig:specific_eturb}. 
We find that the characteristic vortex scale using the LSA or Spectral definitions of the turbulent kinetic energy is of a few kilometers over the entire gain region.
This characteristic vortex scale is both smaller than the size of the averaging cube used ($L=20$~km) in the LSA method and $1/k_{\mathrm{min}}$ used in the Spectral method ($1/k_{\mathrm{min}}=40$~km). \citet{Endeve_2012} use this same quantity to also justify their choice of $k_{\mathrm{min}}$. 
This is an indicator that the LSA and Spectral methods are sufficiently capturing the turbulence. 
In contrast, using the SSA to compute the specific turbulent kinetic energy we obtain a characteristic vortex scale which is generally larger than when using the LSA or Spectral method. 
From Figure \ref{fig:vortexscale} we see that when using the SSA method, the characteristic vortex scale can be as large as $\ell \sim 100\;{\rm km}$.
A vortex of this size more closely resembles the size of the convective B\'enard cells which extend across the gain region rather than local turbulence. 

In summary, we find that different definitions of the turbulent kinetic energy lead to similar, but distinct, measurements of this quantity relative to the total kinetic energy in the simulations.
Using the SSA method we observe a growing amount of turbulent kinetic energy relative to the total kinetic energy from the onset of convection, whereas the LSA and Spectral methods find a somewhat constant (and perhaps even diminishing) amount of turbulence over the same period. 
The SSA method clearly distinguishes between successful and failing supernovae, whereas the LSA and Spectral method do not show a distinction.
Further, maps of the specific turbulent kinetic energy using the SSA method are seen to be dominated by the convective plumes, whereas maps made using the LSA method more closely match the enstrophy, another metric of turbulence.


\section{The Turbulent and Effective Adiabatic Index} \label{sec:TurbulentAdiabticIndex}

\begin{figure*}
    \centering
    \includegraphics[width=\textwidth, trim=0 0 0 0, clip]{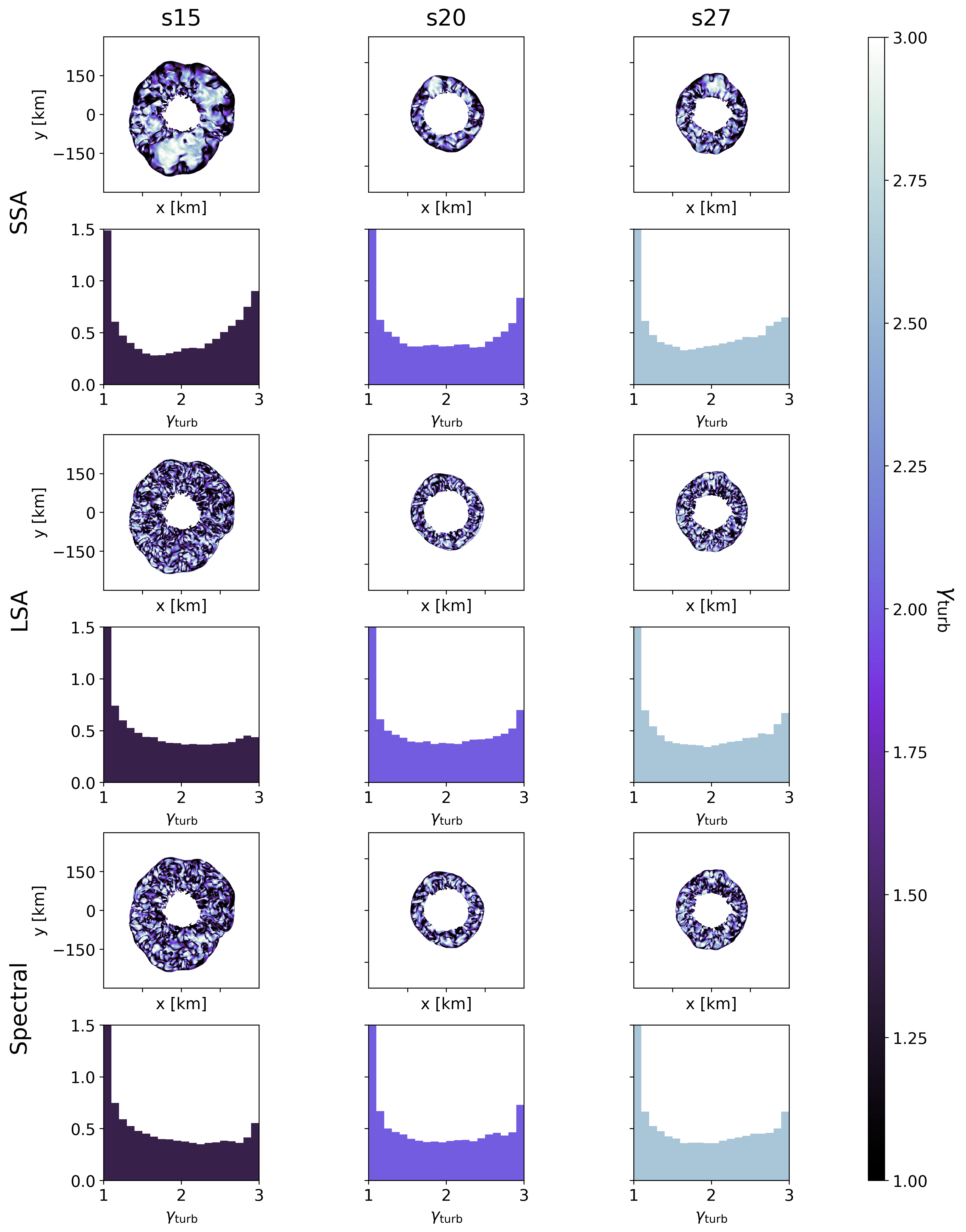}
    \caption{
    Top: Maps in the $xy$-plane and histograms of the turbulent adiabatic index $\gamma_{\turb}$ in the gain region for the s15, s20, and s27 simulations using SSA (top), LSA (middle), and the Spectral method (bottom row).
    \label{fig:gammagrid}
    }
\end{figure*}

\begin{table*}
\centering
\caption{
    Average thermal, turbulent, and effective adiabatic indices, and difference ($\Delta$) of $\overline{\gamma}_{\eff}$ relative to $\overline{\gamma}_{\mathrm{thermal}}$. 
    \label{tab:allgammas}
}
\begin{tabular}{l c ccc ccc ccc}
\hline \hline
Simulation & 
$\overline\gamma_{\thermal}$ & 
\multicolumn{3}{c}{$\overline{\gamma}_{\turb}$} & 
\multicolumn{3}{c}{$\overline\gamma_{\eff}$} & 
\multicolumn{3}{c}{$\Delta$} \\
  & &  SSA & LSA & Spectral & SSA & LSA & Spectral & SSA & LSA & Spectral \\
 \hline
  s15 & 1.4271 & 1.9985 & 1.8171 &1.8203 & 1.4961 & 1.4481 &1.4522 &0.0483  & 0.0147 & 0.0176    \\
 s20 & 1.4364 & 1.9334 & 1.9084 & 1.8986 & 1.4777  & 1.4635 & 1.4616 & 0.0288 & 0.0189 & 0.0175 \\
 s27 & 1.4326 & 1.9362 & 1.9044 & 1.8998 & 1.4933 & 1.4630 & 1.4623 &0.0424 & 0.0212 & 0.0207\\ 
 \hline
 \end{tabular}
\tablecomments{$\Delta = (\overline{\gamma}_{\thermal} - \overline{\gamma}_{\eff, \mathrm{X}})/\overline \gamma_{\thermal}$ where X represents SSA, LSA, or Spectral, respectively.
}
\end{table*}

We can now address the question of how the turbulence alters the fluid properties, and how one's conclusions about the effect of turbulence depends upon the definition.
The quantities we focus upon are the adiabatic indices: the turbulent adiabatic index and the effective adiabatic index. 
The turbulent adiabatic tensor is defined as 
\begin{equation}
(\gamma_{\turb})_{ij} = 1 + \frac{R_{ij}}{\rho\, e_{\turb}},
\label{eq:gammaturb}
\end{equation}
where $R_{ij}$ the $ij^{th}$ element of the Reynolds stress tensor.
For the LSA and SSA we define $R_{ij}$ by $R_{ij} = \rho\,\left( v_i - \langle \boldsymbol{v} \rangle_i\right)\,\left( v_j - \langle \boldsymbol{v} \rangle_j\right)$, where, again, $\langle \boldsymbol{v} \rangle$ is the expectation value of the fluid velocity, and $\rho\,e_{\turb}$ is the turbulent kinetic energy density. 
As discussed in \S\ref{sec:enstrophycomparison}, we can recover the turbulent velocity field, and thus the Reynolds stress tensor, for the Spectral method as follows. 
We compute $R_{ij}$ by Fourier transforming the velocity $\boldsymbol{v}$, setting the amplitude of all Fourier modes to zero for all wavenumbers $k \leq k_{\mathrm{min}}$, then transforming back to real space to produce the velocity field $\boldsymbol{u}$. We then define $R_{ij}$ as $R_{ij} = \rho\, u_i \,u_j$.
Due to its effect upon the shock radius, the most important component of the tensor is the $ij = rr$ element; therefore we will refer to $(\gamma_{\turb})_{rr}$ as the turbulent adiabatic index $\gamma_{\turb}$.


For well-developed isotropic turbulence, one expects each diagonal element (such as $rr$) of the tensor $(\gamma_{\turb})_{ij}$ to have a triangular distribution. If the turbulence is anisotropic such that the distribution of $(\gamma_{\turb})_{rr}$ is the same as the distribution of the sum $(\gamma_{\turb})_{\theta\theta} + (\gamma_{\turb})_{\phi\phi}$ --- as has been found in the simulations reported in \citet{2013ApJ...765..110D} --- we would expect $\gamma_{\turb}$ to have a uniform distribution (see Appendix \ref{sec:app2} for a detailed derivation of these expectations). 
For the case of non-turbulent, bulk flow in the radial direction (either inward or outward), $\gamma_{\turb}$ should have a distribution that peaks toward $\gamma_{\turb} = 3$ when using the SSA method. 
This last expectation arises because, for a location in the bulk flow, $\boldsymbol{v}$ will point approximately radially inwards (for a downflow) or outwards (for a convective plume).
The combination of flows in both directions has the consequence that the average radial velocity $\langle v_r \rangle$ will point inwards (mass is still accreting onto the PNS even if convection is occurring) but $\langle v_r \rangle$ will be smaller than the velocity of the fluid in the downflows.
This has the effect that the Reynolds tensor element $R_{rr}$ will typically be larger than the elements $R_{\theta\theta}$ and $R_{\phi\phi}$.
Given a larger $R_{rr}$, we anticipate that, in the SSA method, our values of $\gamma_{\turb}$ will be pushed upwards toward 3. 
In contrast, for the LSA method, the velocity of a given fluid element relative to the local average velocity would be closer to random than for the SSA, regardless of the relative size of the average fluid velocity $\langle \boldsymbol{v} \rangle$ and the deviation from the average $\boldsymbol{v} -\langle \boldsymbol{v} \rangle$.
Thus we expect the value of $\gamma_{\turb}$ for fluid elements in a bulk flow will not have the same bias toward $\gamma_{\turb} = 3$ when using the LSA method. 
Similarly, when using the Spectral method to compute $\gamma_{\turb}$, we should also find values for $\gamma_{\turb}$ in fluid with bulk flow which are also not biased toward $\gamma_{\turb} = 3$ because this method removes the large scale bulk flow component and retains only the small scale fluid motion. 


In Figure \ref{fig:gammagrid} we show a 2D slice of the turbulent adiabatic index $\gamma_{\turb}$ we obtain in each grid zone, and their histograms at the $t_{\postb} = 150$ ms, using the SSA method (top panels), the LSA method (middle panels), and the Spectral method (bottom panels) for the s15, s20, and s27 simulations. 
The number of bins in every histogram is determined by Sturges's rule \citep{Sturges01031926}
Again, the different methods for defining the turbulence lead to distinct spatial maps.
In the spatial distribution of the LSA and Spectral methods, the structures are of smaller scale, whereas when we use the SSA method, we observe large regions where the adiabatic index has a value $\gamma_{\turb} \geq 2$. 
The histograms of the adiabatic index for the fluid in the gain region are seen to be very similar when using either the LSA or Spectral methods, while the SSA is different.
In particular, the distributions for the SSA method have a more pronounced peak toward $\gamma_{\turb} = 3$ compared with the LSA and Spectral Methods.
The histograms of $\gamma_{\turb}$ are consistent with the idea that the fluid in the gain region can be divided into three components: the fluid that has well-developed, isotropic turbulence, the fluid that has anisotropic turbulence, and the fluid that is only moderately turbulent and has a bulk flow. 
Each of these components of the fluid in the gain region result in different pieces of the distribution of $\gamma_{\turb}$ in the gain region; e.g.\ a triangular distribution for $\gamma_{\turb}$ in the case of isotropic turbulence or a uniform distribution for $\gamma_{\turb}$ for anisotropic turbulence.
For the moderately turbulent component the distribution for $\gamma_{\turb}$ is peaked toward $\gamma_{\turb} = 3$ when using the SSA; for the LSA or Spectral methods the distribution of $\gamma_{\turb}$ will be more similar to the distribution of the other two fluid components. 

\begin{figure*}
    \centering
    \includegraphics[width=\textwidth]{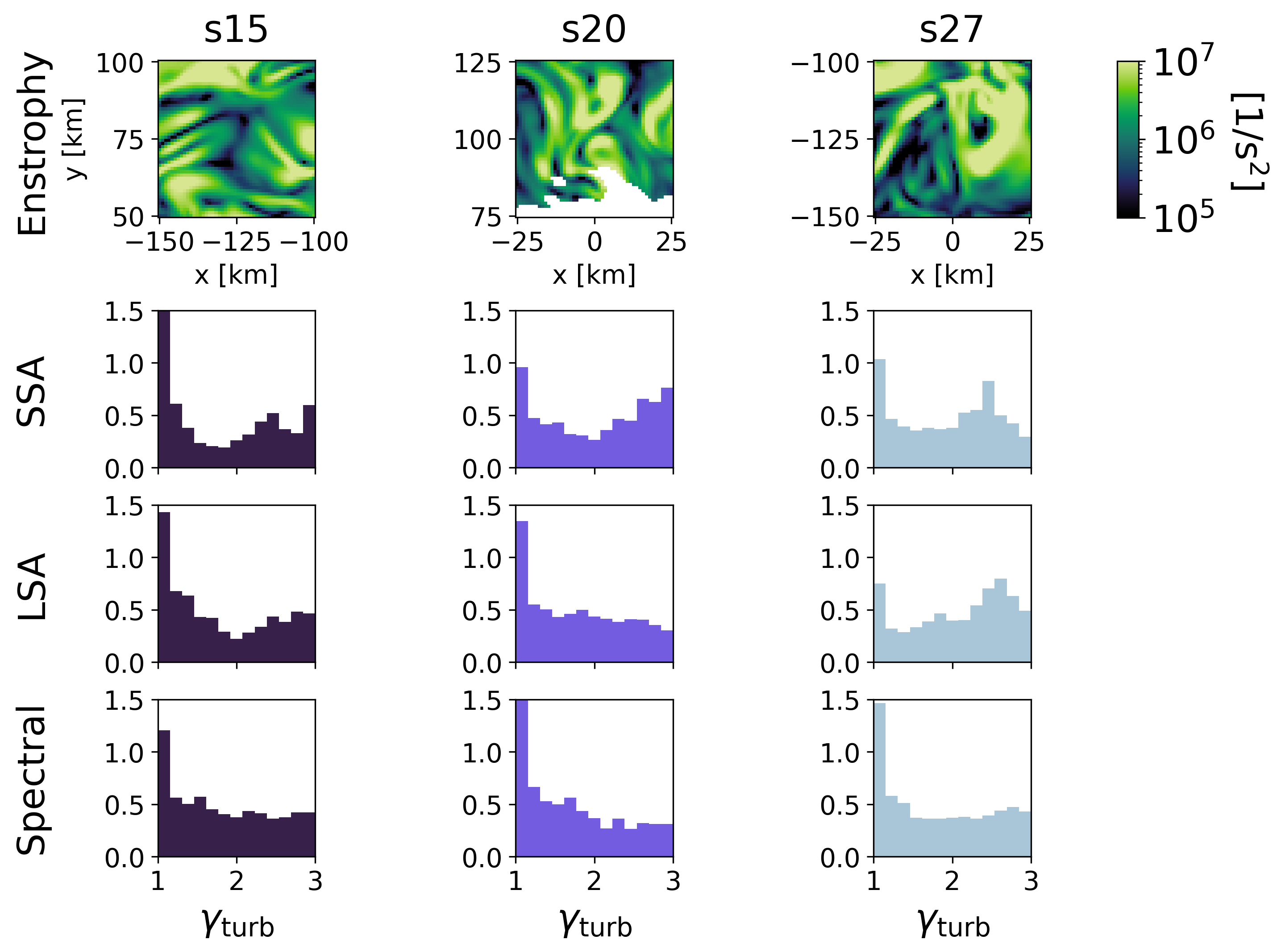}
    \caption{Enstrophy map in the $xy$-plane (top row) and distribution of $\gamma_{\turb}$ for a high-enstrophy patch in the gain region. The $\gamma_{\turb}$ are computed using the SSA method (second row), the LSA method (third row), and the Spectral method (bottom row).
    \label{fig:highgamma}
    }
\end{figure*}
\begin{figure*}
    \centering
    \includegraphics[width=\textwidth]{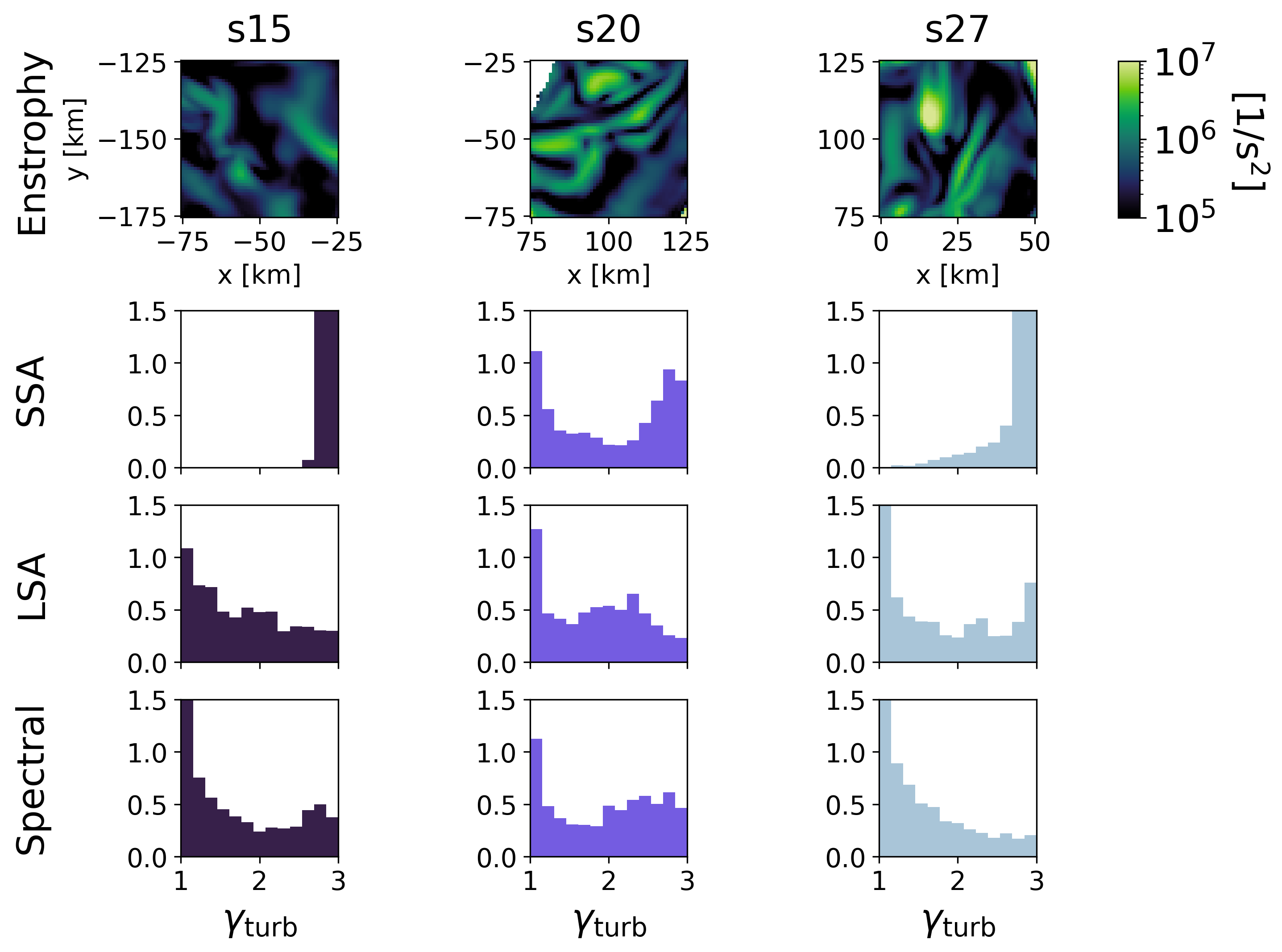}
    \caption{Enstrophy map in the $xy$-plane (top row) and distribution of $\gamma_{\turb}$ for a low-enstrophy patch in the gain region. The $\gamma_{\turb}$ are computed using the SSA method (second row), the LSA method (third row), and the Spectral method (bottom row).
    \label{fig:lowgamma}
    }
\end{figure*}

The different components of the fluid can be seen more clearly if we select small patches of the gain region. We select two 50~km square patches in each simulation, one with high enstrophy and the other with low enstrophy.
The patches we select and the histograms of $\gamma_{\turb}$ for these patches are shown in Figures \ref{fig:highgamma} and \ref{fig:lowgamma}. 
First, the high enstrophy regions in Figure \ref{fig:highgamma} demonstrate that all the methods for measuring the turbulence give similar distributions of $\gamma_{\turb}$ in regions of high enstrophy albeit with a tendency toward larger $\gamma_{\turb}$ from the SSA method. 
Conversely, Figure \ref{fig:lowgamma} demonstrates that in regions of low enstrophy, the different methods yield very different distributions of $\gamma_{\turb}$.

\begin{figure*}
\centering
\includegraphics[width=\textwidth]{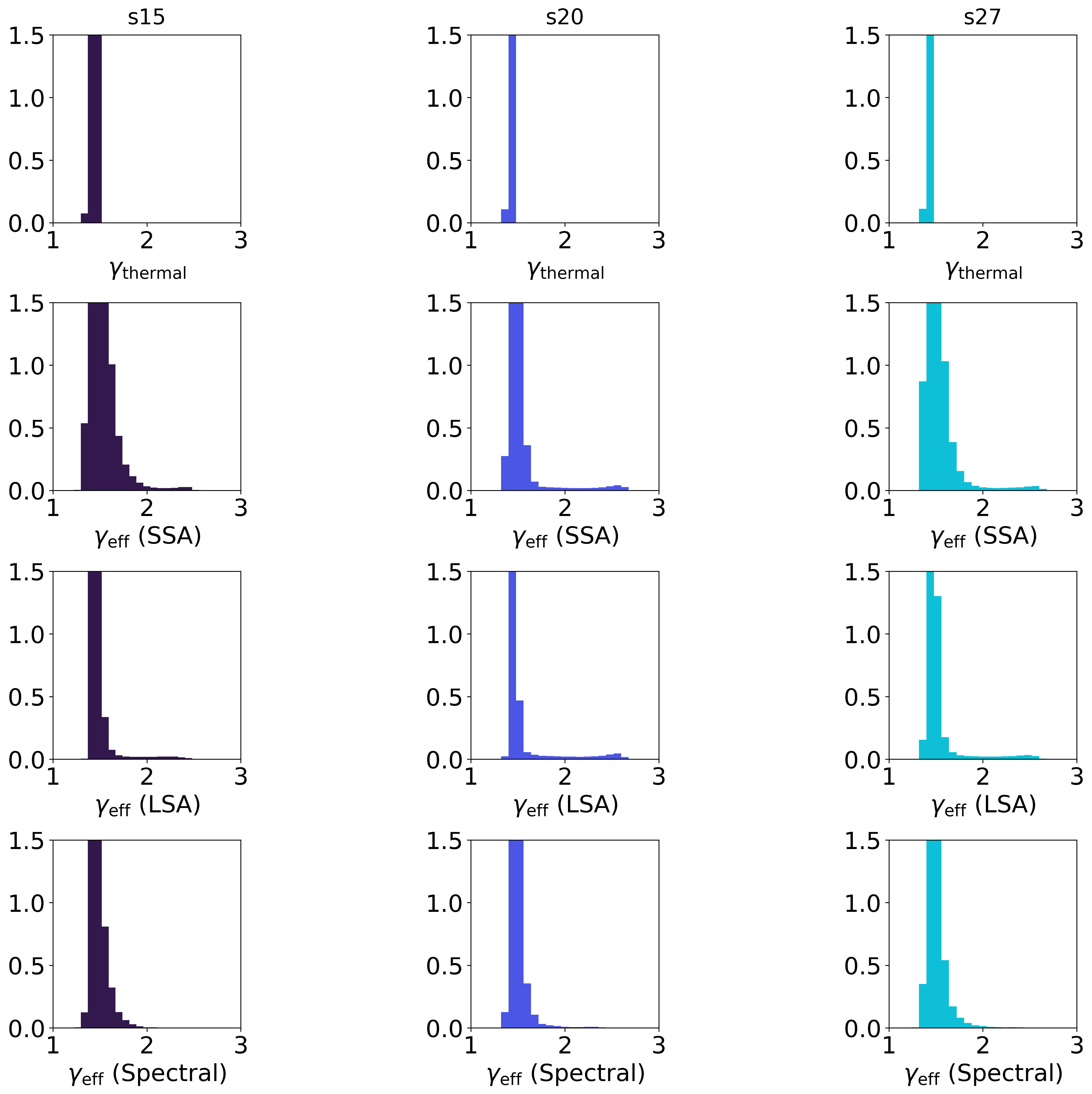}
\caption{Distribution of the thermal adiabatic index $\gamma_{\thermal}$ (top) and the effective adiabatic index $\gamma_{\eff}$ using the SSA (2nd row), LSA (3rd row), and Spectral method (bottom).
    \label{fig:gammaeff}
}
\end{figure*}

The effect of turbulence upon the fluid and explosion, if any, needs to be compared with other sources of energy and pressure in the fluid. 
For this purpose, we can define an effective adiabatic index to be:
\begin{equation}
    \gamma_{\eff} = 1+ \frac{R_{rr} + P_{\thermal}}{E_{\turb}+E_{\thermal}} = \left(\frac{\alpha}{1+\alpha}\right)\,\gamma_{\turb} + \left(\frac{1}{1+\alpha}\right)\,\gamma_{\thermal},
    \label{eq:gammaeff}
\end{equation}
where $E_{\thermal}=\rho\,e_{\thermal}$ and $P_{\thermal}$ represent the specific thermal energy and pressure, $\gamma_{\thermal} = 1 + P_{\thermal}/(\rho\,e_{\thermal})$, and $\alpha = e_{\turb}/e_{\thermal}$. 
In Figure \ref{fig:gammaeff} we show histograms of $\gamma_{\thermal}$ and $\gamma_{\eff}$ for the gain region from the same snapshots at 150~ms post bounce of the exploding simulations using the three different turbulence calculation methods.  
The thermal adiabatic index is approximately $\gamma_{\thermal} = 1.43$ for all fluid within the gain region at this time. 
When we add the contribution from turbulence and compute $\gamma_{\eff}$, the peak at 
1.43 
is retained but its width increases. 
When we use the LSA or Spectral methods, the broadening is smaller than when we use the SSA method. 
The addition of the turbulence shifts the average value of  $\gamma_{\eff}$ upwards, and the average $\gamma_{\eff}$ for these snapshots are shown in Table \ref{tab:allgammas}.
Whatever method we use to define the turbulence, the shift of $\gamma_{\eff}$ from $\gamma_{\thermal}$ is not large; for the successful explosions, $\gamma_{\eff}$ is larger than $\gamma_{\thermal}$ by, at most, $5\%$, as seen in Table \ref{tab:allgammas}. 
Notably the largest shifts come from the SSA method, while similar but smaller shifts come from the LSA and Spectral methods. 
This further confirms the notion that the methods can suggest different amounts of importance of turbulence relative to other physics.

\begin{figure}
\centering
\includegraphics[width=0.45\textwidth]{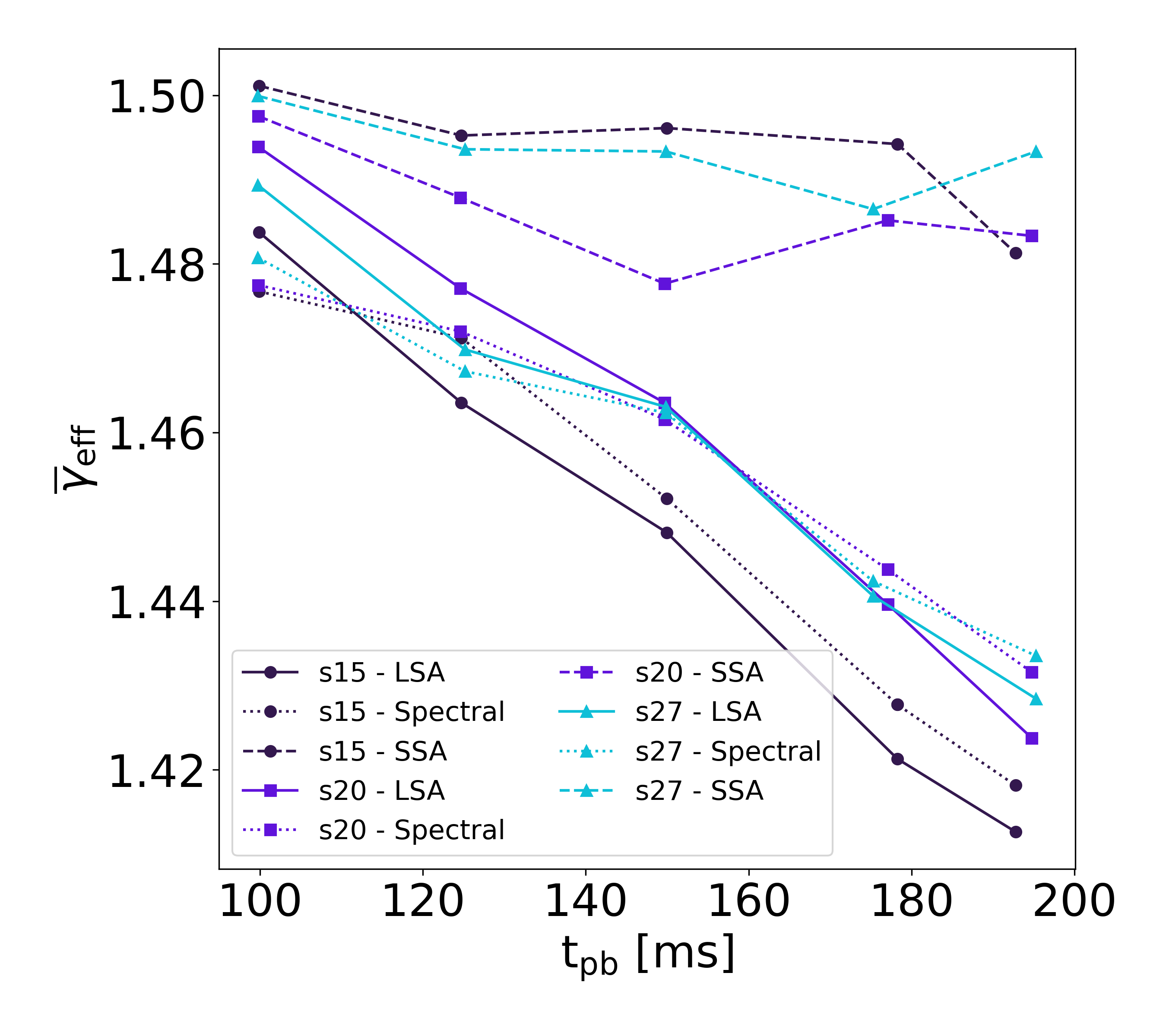}
    \caption{Average effective adiabatic index $\overline{\gamma}_{\eff}$ in the gain region as a function of post-bounce time using the LSA (solid line), SSA (dashed line), and Spectral method (dotted line) for the s15, s20, and s27 simulations.
    \label{fig:gammaeffovert}
    }
\end{figure}

\begin{figure}
\centering
\includegraphics[width=0.45\textwidth]{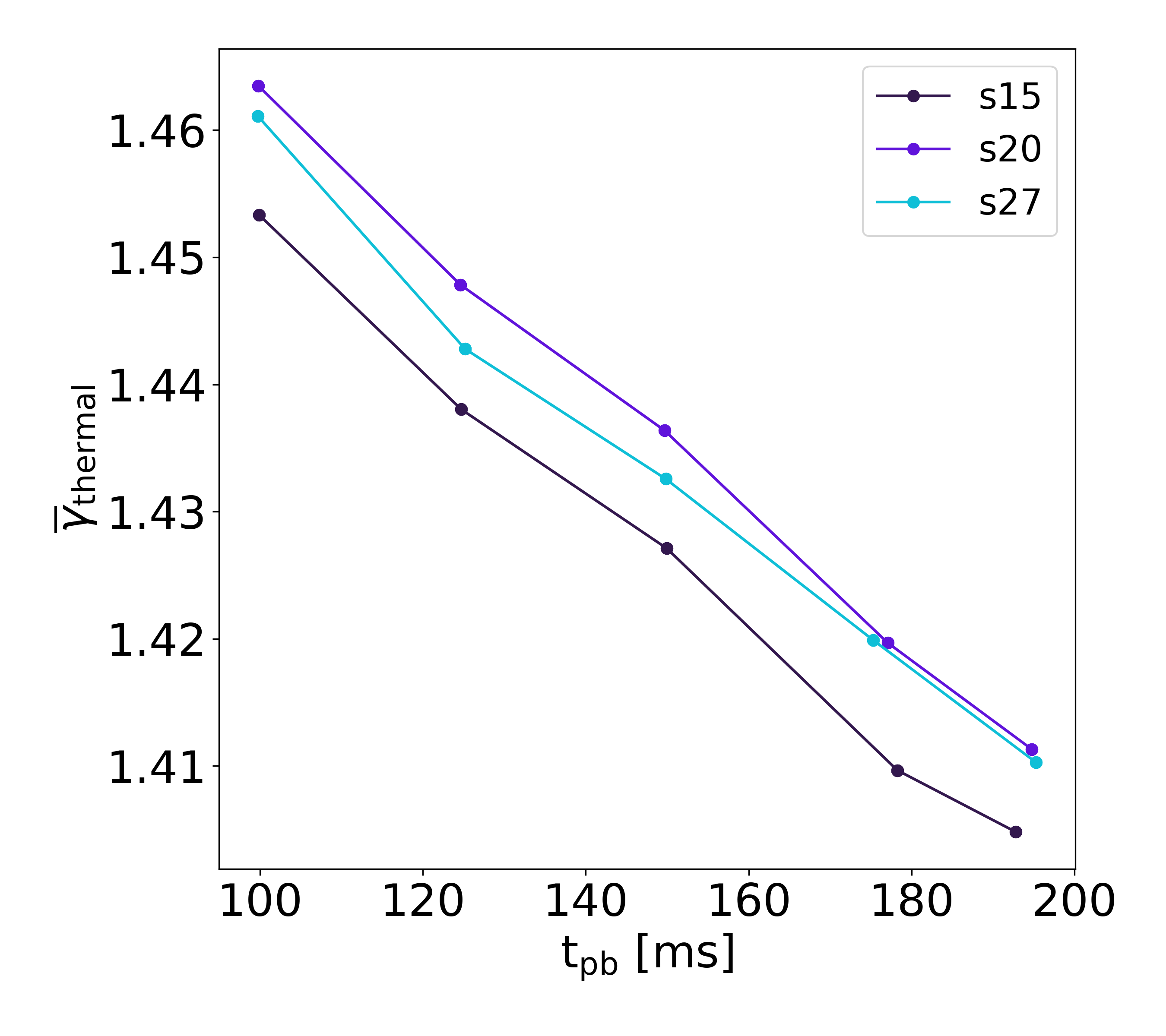}
    \caption{Average thermal adaibatic index $\overline{\gamma}_{\thermal}$ in the gain region as a function of post-bounce time for the s15, s20, s27 simulations.
    \label{fig:gamma_thermal_over_t}
    }
\end{figure}

Finally, in Figures \ref{fig:gammaeffovert} and \ref{fig:gamma_thermal_over_t}  we show the evolution of the average thermal and effective adiabatic indices for the entire gain region as a function of the post-bounce time. 
As the fluid in the gain region is heated, its thermal adiabatic index gradually decreases due to a greater proportion of relativistic material to non-relativistic material.
The decrease is not large; only 3\% over the 100 ms period shown. 
The evolution of $\gamma_{\eff}$ over this same period, on the other hand, depends upon the method used to define turbulence, as shown in Figure \ref{fig:gammaeffovert}. 
In all three successful explosions, the SSA method yields an average $\gamma_{\eff}$ which is larger than for the other two definitions of turbulence, and remains consistently between 1.48 and 1.5.
This behavior is just as we saw at the single time point above, but we can now generalize to more of the simulation as a function of post-bounce time.
In contrast, the LSA and Spectral methods for computing the turbulence yield average $\gamma_{\eff}$ which gradually decrease with time, although by only $\sim 4\%$.

\section{Summary, Conclusions and Discussion} \label{sec:conclusions}

In this paper, we have investigated the question of how different definitions of the turbulence alters one's conclusions about the effect of turbulence in supernova simulations.
We have undertaken four simulations using four progenitors and used three different methods for computing the turbulence.
We find that how the turbulence is defined alters the amount of measured turbulent kinetic energy in the gain region.
The Spherical Spatial Average method consistently yields values for the turbulent kinetic energy which are larger than the values obtained with the Local Spatial Average and Spectral Methods. 
The differences are only partly explained away by the choices one has to make for the averaging volume scale in the LSA method and the minimum wavenumber in the Spectral Method.
The remaining difference is indicative of the fundamental differences of the definitions. 
These fundamental differences are seen most clearly in Figure \ref{fig:TurbE}, which shows how the ratio of the turbulent kinetic energy to total kinetic energy in the gain region evolves with time.
The SSA method finds a consistently larger ratio of the energies for successful explosions and the ratio tends to grow as a function of post-bounce time. 
Using the SSA one would conclude that the presence of turbulence correlates with the success of the explosion. 
In contrast, the ratio of energies calculated using the LSA and Spectral methods are similar for both successful \emph{and} unsuccessful explosions.
Using these methods one would conclude that the presence of turbulence does not correlate with the success of the supernova. 
This divergence of conclusions indicates that the different methods of measuring the turbulence are actually not measuring the same thing. 

To determine which method, if any, is a better measure of the turbulence we compared maps of the specific turbulent kinetic energy with maps for the enstrophy, a quantity that is often associated with the presence of turbulence. 
The comparison revealed that the LSA and Spectral methods produce maps which are more similar to the enstrophy than the SSA method. 

Finally, we examined the effect of turbulence upon the fluid by computing the turbulent and effective adiabatic indices.
Maps of the turbulent adiabatic index calculated using the SSA show larger coherent regions with $\gamma_{\text{turb}} > 2$ compared to those produced by the LSA and Spectral methods.
These large regions of high $\gamma_{\turb}$ alter the distributions of this adiabatic index, producing a peak in the distribution toward the maximum value $\gamma_{\turb} = 3$.
The net effect is to shift the average turbulent adiabatic index in the gain region upward by a larger amount when using the SSA definition compared to the shift when using either of the other two definitions. 

The contribution of turbulence to the explosion can be measured by computing the effective adiabatic index $\gamma_{\eff}$.
All definitions of turbulence shift the average $\gamma_{\eff}$ upwards.
However, the shift is more pronounced when we use the SSA method compared to the LSA and Spectral definitions.
Furthermore, the average $\gamma_{\eff}$ in the gain region remains approximately constant when we use the SSA. In contrast, when we use the LSA or Spectral definitions, the average $\gamma_{\eff}$ decreases with time. These findings once again demonstrate that the method selected for defining the turbulence changes the interpretation of the impact of turbulence as a function of post-bounce time. 

When assessing the impact of turbulence on the evolution of CCSNe, it is necessary to understand the strengths and limitations of the method used to measure it.
We conclude from our study that the SSA method for defining the turbulence is not as accurate of a measure of this quantity as the LSA and Spectral methods.
The LSA and Spectral methods are more consistent with each other \emph{and} with the distribution of enstrophy in the simulations.
The SSA yields a turbulent kinetic energy which is dominated by the upflowing convective plumes, not actual turbulence.
However, the LSA and Spectral definition are not free of systematic problems.
For example, using the LSA for the fluid in the vicinity of counterflows means there is a great deal of cancellation of momentum within the averaging volume which tends to yield low average velocities, and thus larger contributions to the turbulent kinetic energy.
Further, both methods require one to define an additional parameter, and we have found that the measure of the turbulent kinetic energy is not independent of this choice.
Thus the LSA and Spectral Methods suffer from an element of subjectivity that the SSA does not share.
Nevertheless, the systematic issues of the LSA and Spectral definitions appear to not have as great an effect upon turbulent kinetic energy as convection has upon the SSA. 

While the LSA and Spectral methods correlate more closely to the often-used turbulence tool of enstrophy, the quantity measured by the SSA method is clearly useful because it correlates with the success of the explosion. 
Since the turbulent kinetic energy computed using the SSA is dominated by convective plumes, the correlation suggests that the transition to explosion is because the kinetic energy in the gain region is increasingly directed toward countering the ram pressure of the infalling material. 
Since the turbulent kinetic energy computed using the SSA is dominated by convective plumes, the correlation suggests that the SSA is more likely tracking convection than turbulence.
While there are a multitude of mechanisms, such as neutrino heating and deleptonization, contributing to the convection, the connection between SSA and explosion can be more accurately attributed to the necessity for convection for a supernova to explode

\begin{acknowledgments}
The work at NC State was supported by United States Department of Energy, Office of Science, Office of Nuclear Physics (award number DE-FG02-02ER41216).
The work of Michael Redle was partially funded by German Research Foundation (DFG) Research Unit FOR5409 ``Structure-Preserving Numerical Methods for Bulk- and Interface-Coupling of Heterogeneous Models (SNuBIC)" (grant \#463312734). 
This work used Bridges-2 at Pittsburgh Supercomputing Center through allocation PHY200074 from the Advanced Cyberinfrastructure Coordination Ecosystem: Services \& Support (ACCESS) program, which is supported by National Science Foundation grants \#2138259, \#2138286, \#2138307, \#2137603, and \#2138296. This work also used the Extreme Science and Engineering Discovery Environment (XSEDE), which is supported by National Science Foundation grant number ACI-1548562. Specifically, it used the Bridges-2 system, which is supported by NSF award number ACI-1928147, at the Pittsburgh Supercomputing Center (PSC).
The authors also acknowledge the Texas Advanced Computing Center (TACC) at The University of Texas at Austin for providing computational resources that have contributed to the research results reported within this paper. URL: http://www.tacc.utexas.edu
\end{acknowledgments}

\software{ELEPHANT \citep{Liebendoerfer.IDSA:2009,kappeli2011fish}, astropy \citep{2013A&A...558A..33A,2018AJ....156..123A,2022ApJ...935..167A}, Matplotlib \citep{Hunter:2007}, NumPy \citep{harris2020array}
}
          

\appendix

\section{The Distribution of the Turbulent Adiabatic Index in Well-Developed Turbulence}
\label{sec:app2}

The elements of the adiabatic index tensor for turbulence are defined to be
\begin{equation}
(\gamma_{\turb})_{ij} = 1 + \frac{R_{ij}}{E_{\turb}}
\end{equation}
with $R_{ij}$ the $ij$'th element of the Reynolds stress tensor given by $R_{ij} = \rho\,\left( v_i - \langle \vec{v} \rangle_i\right)\,\left( v_j - \langle \vec{v} \rangle_j\right)$, $\langle \vec{v} \rangle$ is the expectation value of the fluid velocity, and $E_{\turb}$ is the turbulent kinetic energy $E_{\turb} = \rho\, |\vec{v} - \langle \vec{v} \rangle|^2/2$. For our derivation of the distribution of the adiabatic index in well-developed turbulence, it does not matter how $\langle \vec{v} \rangle$ is defined. The diagonal elements of the Reynolds stress tensor are non-negative, and from the definition of the Reynolds stress tensor, we see that 
\begin{equation}
2\,E_{\turb} = \,\sum_i R_{ii}.
\end{equation}
Thus the three diagonal components of the Reynold's stress tensor are not independent, they are constrained, and the constraint defines a simplex in the space of $R_{rr}$, $R_{\theta\theta}$ and $R_{\phi\phi}$ (using spherical coordinate labels for the diagonal elements of $R_{ij}$). 

\subsection{Isotropic Turbulence}
In isotropic turbulence the values for the diagonal elements of the Reynold's stress tensor at a given location in the fluid may be regarded as random and equally likely i.e. each diagonal element of the stress tensor has the same probability distribution. This means that, for a fixed total energy $E_{\turb}$,  the infinitesimal probability $dP(R_{rr},R_{\theta\theta},R_{\phi\phi})$ of finding the diagonal elements of the stress tensor in the ranges $R_{rr}$ to $R_{rr} + dR_{rr}$, $R_{\theta\theta}$ to $R_{\theta\theta} + dR_{\theta\theta}$, and $R_{\phi\phi}$ to $R_{\phi\phi} + dR_{\phi\phi}$, 
is uniform across the surface of the simplex and so proportional to the surface element $dA = \delta(2\,E_{\turb} - R_{rr} - R_{\theta\theta} - R_{\phi\phi})\, dR_{rr}\, dR_{\theta\theta}\, dR_{\phi\phi}$ of the simplex. Thus we can write the probability of a given set of diagonal elements as
\begin{equation}
dP(R_{rr},R_{\theta\theta},R_{\phi\phi}) \propto \delta(2\,E_{\turb} - R_{rr} - R_{\theta\theta} - R_{\phi\phi})\, dR_{rr}\, dR_{\theta\theta}\, dR_{\phi\phi}.
\end{equation}
To find the distribution of any one diagonal element, e.g. $R_{rr}$, we integrate this probability distribution over the other two diagonal elements i.e.
\begin{equation}
\frac{dP(R_{rr})}{dR_{rr} } = C \int_{0}^{2\,E_{\turb}}\,\int_{0}^{2\,E_{\turb}}\,\delta(2\,E_{\turb} - R_{rr} - R_{\theta\theta} - R_{\phi\phi})\, dR_{\theta\theta}\,dR_{\phi\phi}.
\end{equation}
where $C$ is a normalization constant. To evaluate the integral we first insert two Heaviside step functions so that we can change the limits of the integration for one variable, e.g. $R_{\phi\phi}$, and write 
\begin{equation}
\frac{dP(R_{rr})}{dR_{rr} } = C \int_{0}^{2\,E_{\turb}}\,\int_{-\infty}^{\infty}\,\delta(2\,E_{\turb} - R_{rr} - R_{\theta\theta} - R_{\phi\phi})\,\Theta(2\,E_{\turb} - R_{\phi\phi})\,\Theta(R_{\phi\phi})\,dR_{\theta\theta}\,dR_{\phi\phi}.
\end{equation}
From the integral over $R_{\phi\phi}$ we obtain
\begin{equation}
\frac{dP(R_{rr})}{dR_{rr} } = C \int_{0}^{2\,E_{\turb}}\,\Theta(2\,E_{\turb}-R_{rr}-R_{\theta\theta})\,\Theta(R_{rr}+R_{\theta\theta})\,dR_{\theta\theta}
\end{equation}
and now we can remove the step functions by changing the limits of the remaining integral over $R_{\theta\theta}$ to give 
\begin{equation}
\frac{dP(R_{rr})}{dR_{rr} } = C \int_{0}^{2\,E_{\turb}-R_{rr}}\,dR_{\theta\theta}
\end{equation}
The integral is now trivial and after normalizing we find
\begin{equation}
\frac{dP(R_{rr})}{dR_{rr}} = \frac{1}{E_{\turb}}\,\left(1-\frac{R_{rr}}{2\,E_{\turb}}\right)
\end{equation}
This is a triangular distribution for $R_{rr}$. Now we know the distribution of $R_{rr}$ we see the distribution for $\gamma_{rr}$ (or any other diagonal element of the tensor) is also a triangular distribution  
\begin{equation}
\frac{dP(\gamma_{rr})}{d\gamma_{rr}} = \frac{(3-\gamma_{rr})}{2}
\end{equation}
with a range $1 \leq \gamma_{rr} \leq 3$. The average adiabatic index is $\langle \gamma_{rr} \rangle = 5/3$ which matches the well known result of the adiabatic index of isotropic turbulence \citep{2018JPhG...45e3003R}.

\subsection{Anisotropic Turbulence}
For anisotropic turbulence, one of the diagonal components of the Reynold's stress has a distribution different from the other two. In many models of anisotropic turbulence e.g. \cite{2013ApJ...771...52M}, the distribution of the biased component is equal to the distribution of the sum of the other two components. Defining $R_{tt} = R_{\theta\theta} + R_{\phi\phi}$, the probability 
$dP(R_{rr},R_{tt})$ of finding the elements of the stress tensor in the ranges $R_{rr}$ to $R_{rr} + dR_{rr}$ and $R_{tt}$ to $R_{tt} + dR_{tt}$, is uniform across the line $2\,E_{\turb} = R_{rr} + R_{tt}$ and so proportional to the line element $d\ell = \delta(2\,E_{\turb} - R_{rr} - R_{tt})\, dR_{rr}\, dR_{tt}$. We can proceed exactly as before to integrate over $R_{tt}$ to determine the distribution of $R_{rr}$. We find
\begin{equation}
\frac{dP(R_{rr})}{dR_{rr}} = \frac{1}{2\,E_{\turb}}\,\Theta(R_{rr})\,\Theta(2\,E_{\turb} - R_{rr})
\end{equation}
which is a uniform distribution. Thus $\gamma_{rr} = 1 + R_{rr} / E_{\turb}$ is also uniform and equal to
\begin{equation}
\frac{dP(\gamma_{rr})}{d\gamma_{rr}} = \frac{1}{2}\,\Theta(\gamma_{rr}-1)\,\Theta(3-\gamma_{rr}).
\end{equation}
From this result we recover the result that for anisotropic turbulence, the average adiabatic index is $\langle \gamma_{rr} \rangle = 2$ \citep{2018JPhG...45e3003R}.

\bibliographystyle{aasjournalv7}
\bibliography{references}

\end{document}